\documentclass[aps,prb,amsmath,amssymb,twocolumn,superscriptaddress,letterpaper,english]{revtex4}
\usepackage{times}
\usepackage{amsfonts}
\usepackage{mathrsfs}
\usepackage{graphicx}
\usepackage{dcolumn}
\usepackage{bm}
\usepackage{color}

\usepackage[colorlinks,bookmarks=false,citecolor=blue,linkcolor=red,urlcolor=blue]{hyperref}

\usepackage{url}

\usepackage{multirow}

\def\be{\begin{equation}}       \def\ee{\end{equation}}
\def\bea{\begin{eqnarray}}      \def\eea{\end{eqnarray}}

\begin{document}
\title{Harmonic Fingerprint of Unconventional Superconductivity in Twisted Bilayer Graphene}

\author{Xianxin Wu}
\email{xianxinwu@gmail.com}
\affiliation{Institut f\"ur Theoretische Physik und Astrophysik,
  Julius-Maximilians-Universit\"at W\"urzburg, 97074 W\"urzburg,
  Germany}
\author{Werner Hanke}
\affiliation{Institut f\"ur Theoretische Physik und Astrophysik,
  Julius-Maximilians-Universit\"at W\"urzburg, 97074 W\"urzburg,
  Germany}

\author{Mario Fink}
\affiliation{Institut f\"ur Theoretische Physik und Astrophysik,
  Julius-Maximilians-Universit\"at W\"urzburg, 97074 W\"urzburg,
  Germany}

\author{Michael Klett}
\affiliation{Institut f\"ur Theoretische Physik und Astrophysik,
  Julius-Maximilians-Universit\"at W\"urzburg, 97074 W\"urzburg,
  Germany}

\author{Ronny Thomale}
\email{ronny.thomale@physik.uni-wuerzburg.de}
\affiliation{Institut f\"ur Theoretische Physik und Astrophysik,
  Julius-Maximilians-Universit\"at W\"urzburg, 97074 W\"urzburg,
  Germany}

\date{\today}

\begin{abstract}
Microscopic details such as interactions and Fermiology determine the structure of superconducting pairing beyond the spatial symmetry classification along irreducible point group representations. From the effective pairing vertex, the pairing wave function related to superconducting order unfolds in its orbital-resolved Fourier profile which we call the harmonic fingerprint (HFP). The HFP allows to formulate a concise connection between microsopic parameter changes and their impact on superconductivity. From a random phase approximation analysis of twisted bilayer graphene (TBG) involving $d+id$, $s_{\pm}$, and $f$-wave order, we find that nonlocal interactions, which unavoidably enter the low-energy electronic description of TBG, not only increase the weight of higher lattice harmonics but also have a significant effect on the orbital structure of these pairing states. For gapped unconventional superconducting order such as $s_{\pm}$ and $d+id$, a change in HPF induces enhanced gap anisotropies. Experimental implications to distinguish the different gaps and HPFs are also discussed.
\end{abstract}


\maketitle

\section{Introduction}
The discovery of correlated insulating states and superconductivity (SC) in twisted bilayer graphene (TBG) at the first magic angle has generated enormous recent excitement\cite{Cao2018-1,Cao2018-2}. In apparent similarity to the cuprates, SC emerges upon hole or electron doping away from insulating commensurate fillings\cite{Cao2018-1,Cao2018-2,Yankowitz2019,LiuXB2019}, by tuning carrier density, applying a magnetic field or slightly varying the twist angle\cite{Cao2018-1,Cao2018-2}.

At small twist angles, the corresponding Moir\'{e} pattern gives rise to large unit cells with more than 10000 atoms. In particular, near certain " magic " angles (e.g. at $\sim$1.05$^{\circ}$), four lowest-energy mini-bands with a bandwidth of order 10 meV are formed, well separated from higher-energy bands\cite{Koshino2018PRX}. Due to the substantial suppression of kinetic energy in the narrow bands, Coulomb-interaction effects are strongly enhanced and expected to drive correlated electron phenomena\cite{Cao2018-1,Cao2018-2}.

SC appears, when the Mott-type insulating states, which occur at a filling of charge $\pm$2e per supercell, are slightly hole- or electron-doped. The relatively large $T_c/E_{Fermi}$ ratio, with $T_c\sim1.7$ K and $E_{Fermi}\sim$ 10 meV, which is even larger than in the high-T$_c$ cuprates, in combination with the record-low carrier density of a few $10^{11}$ $e/cm^2$\cite{Cao2018-1}, further points to the possibility in that the TBG-type of SC is of unconventional nature. What makes TBG so challenging for a SC description is, in particular, that it has flat bands, which yield very high density of states (van-Hove type of singularities) at a very low carrier density. Accordingly, a considerable amount of work has so far been devoted to understanding insulating \cite{Po2018PRX,Padhi2018,Dodaro2018,Liu2108PRL,Huang2018,WuXC2018,Pizarro2018,Ochi2018,Isobe2018PRX,Gonzalez2018PRL,Kennes2018,Kang2019PRL}  and superconducting \cite{Po2018PRX,XuCK2018PRL,Liu2108PRL,Huang2018,Isobe2018PRX,Fidrysiak2018,Roy2018,YouYZ2018,Gonzalez2018PRL,Guo2018PRB,Kennes2018,Laksono2018,ZhuGY2018,Lin2018PRB,Tang2018,Kozii2018,LiuZ2018}
phases from an electronic correlation point of view. Typically here, again in some analogy to the cuprates, a standard (multi-orbital) Hubbard model, only including the on-site interactions, has been employed. On the other hand, electron-phonon coupling has also been suggested to play a crucial role in the SC pairing mechanism in TBG\cite{WuFC2018PRL,Lian2018,Peltonen2018,WuFC2018-2}.

The effective electronic band model to describe the low-energy physics in TBG is also still controversial due to a so-called "fragile topology" of the lowest-energy four narrow bands under $D_6$ point-group symmetry. High-energy bands should, in principle, also be included in a "faithful" tight-binding model construction. Such a faithful one is a ten-band model per valley and per spin\cite{Po2018PRX,Zou2018PRB,Song2018,Po2018-2,Carr2019PRR,Carr2019-2,Fang2019}. However, the enormous numerical difficulty in identifying the targeted high-energy bands renders the determination of the parameters extremely difficult in the above ten-band model\cite{Carr2019-2}. Considering that SC is a low-energy phenomenon, the four-band model with lower symmetry $D_3$ appears as a legitimate starting point in TBG\cite{Yuan2018PRB,Kang2018PRX,Koshino2018PRX}. Furthermore, in the presence of SU(2) spin symmetry, the two point groups have the same pairing symmetry classification, i.e., s,p,d,f, etc. In this model, however, the \emph{Wannier orbitals are necessarily rather extended\cite{Yuan2018PRB,Kang2018PRX,Koshino2018PRX} and as a consequence, besides the on-site interaction, non-local interactions become essential.}

As diverse as the current theoretical predictions are, where, in particular, the electronic structure, the SC pairing mechanism and nature of the gap function are still under intense debate, this situation is likely to persist until further experimental evidence confronted with theoretical progress can be accumulated for the superconducting and insulating phases in TBG.

  In this work, we show, that the multi-orbital nature of the TBG system, in combination with the non-local interactions, can have a decisive influence on the gap function and, therefore, should become important for a reliable differentiation of the pairing mechanisms, when compared with experiments.

 In a crystal, the pair wave function can be classified according to its irreducible representation (IR) of the lattice point group. This is, however, neither sufficient to fully specify the precise shape of the superconducting gap nor to uniquely trace its microscopic origin. For instance, an $A_{1g}$ gap can be just a constant in conventional superconductors. But it can also be of $s_{\pm}$ form in iron based superconductors, where hole and electron pockets, located at e.g. $\Gamma$ and $M$ points exhibit a pairing amplitude of opposite sign, relating to a harmonic component of $\Delta_{A_{1g}}({\bf k})\propto (\cos k_x + \cos k_y)$\cite{Stewart2011,Hirschfeld2011,Seo2018}. Even for generically unconventional pairing symmetries such as $B_{1g}$, i.e., $d_{x^2-y^2}$-wave, it was found for hole-doped iron pnictides, that the harmonic composition of $\Delta_{B_{1g}}({\bf k})$ can drastically depend on the  multi-orbital/pocket/sublattice fermiology\cite{Platt2013}. This has e.g. been confirmed by Raman spectroscopy of Bardasis-Schrieffer bound states\cite{Boehm2018}. Similarly, the pronounced low-energy quasiparticle weight observed for the $E_{2u}$ ($p$-wave) pairing candidate Sr$_2$RuO$_4$ might be reconciled with higher-order harmonic pairing contributions\cite{Mackenzie2003,WangQH2013,Scaffidi2015}.

 To this, we develop a systematic pairing wave function analysis of correlation-driven superconductivity in TBG. At the example of the pairing vertex obtained from a random phase approximation (RPA) study for long-range Coulomb interactions and an effective tight-binding model for TBG, we introduce the \emph{notion of a harmonic fingerprint (HFP) of superconducting order}, which corresponds to the harmonic, orbital, sublattice, and spin decomposition of a superconducting pairing state. It establishes a nomenclature in which two superconducting pairings can be compared beyond their mere symmetry classification, and allows to resolve and formulate microscopic parameter trends for superconducting pairing. In the case of TBG, the changes of orbital structure as well as lattice harmonic in the HFP of unconventional pairing are implied by \emph{an enhanced range of electronic interactions, and yield enhanced gap anisotropies which are likely to be traceable in the quasiparticle spectral weight at low energies}.

The outline of our paper is as follows: In Sec. \ref{sectionTB}, we
introduce the tight-binding (TB) model for flat bands and the Fermi surfaces near half filling of valence bands in TBG. In Sec.\ref{pairingRPA}, within the random phase approximation (RPA), we investigate the pairing symmetry slightly away from half filling of valence bands, first with only onsite interactions and then, further with additional nonlocal interactions. In Sec.\ref{realspacepairing}, we classify the pairing states in the two-orbital honeycomb lattice and analyze the harmonic fingerprint, i.e. their HFP of the pairing states from RPA. The nonlocal interactions turn out to have a crucial effect on the orbital structure as well as the lattice harmonics for pairing states, for which we also provide a qualitative understanding using diagrams. In Sec.\ref{conclusion}, we discuss experimental implications and provide the main conclusions of this paper.
In particular, we discuss Scanning Tunneling Microscopy (STM), which is already providing valuable insights, for instance, for the effect of correlations on the flat bands\cite{Choi2018}, as a powerful technique for revealing the HFP of unconventional SC pairing.

\section{Tight binding model and Fermiology}\label{sectionTB}

Presently, there exists an intense discussion on the electronic structure of TBG systems, in particular, on an accurate description of the flat bands. The methodology used ranges from large-scale a-priori density-functional theory (DFT) calculations, empirical tight-binding descriptions\cite{Moon2012,Kang2018PRX}, to low-energy $\bm{k}\cdot\bm{p}$ continuum theories \cite{LopesdosSantos2007,Bistritzer2011}. The latter continuum method, where the states are primarily derived from the vicinity of the Dirac points in each layer, is expected to hold when the twisted angle is small. Clearly, it is a computationally very efficient method and allows twist-angle control of the band structure not constrained by commensurate conditions as in the DFT and tight-binding procedures.

However, at present, a fair statement is that each available methodology has its strengths and weaknesses: for example, recent experiments point to significant deviations compared to the bandwidth extracted from the continuum theory\cite{Tomarken2019}.

\begin{figure*}[tb]
\centerline{\includegraphics[width=2.2\columnwidth]{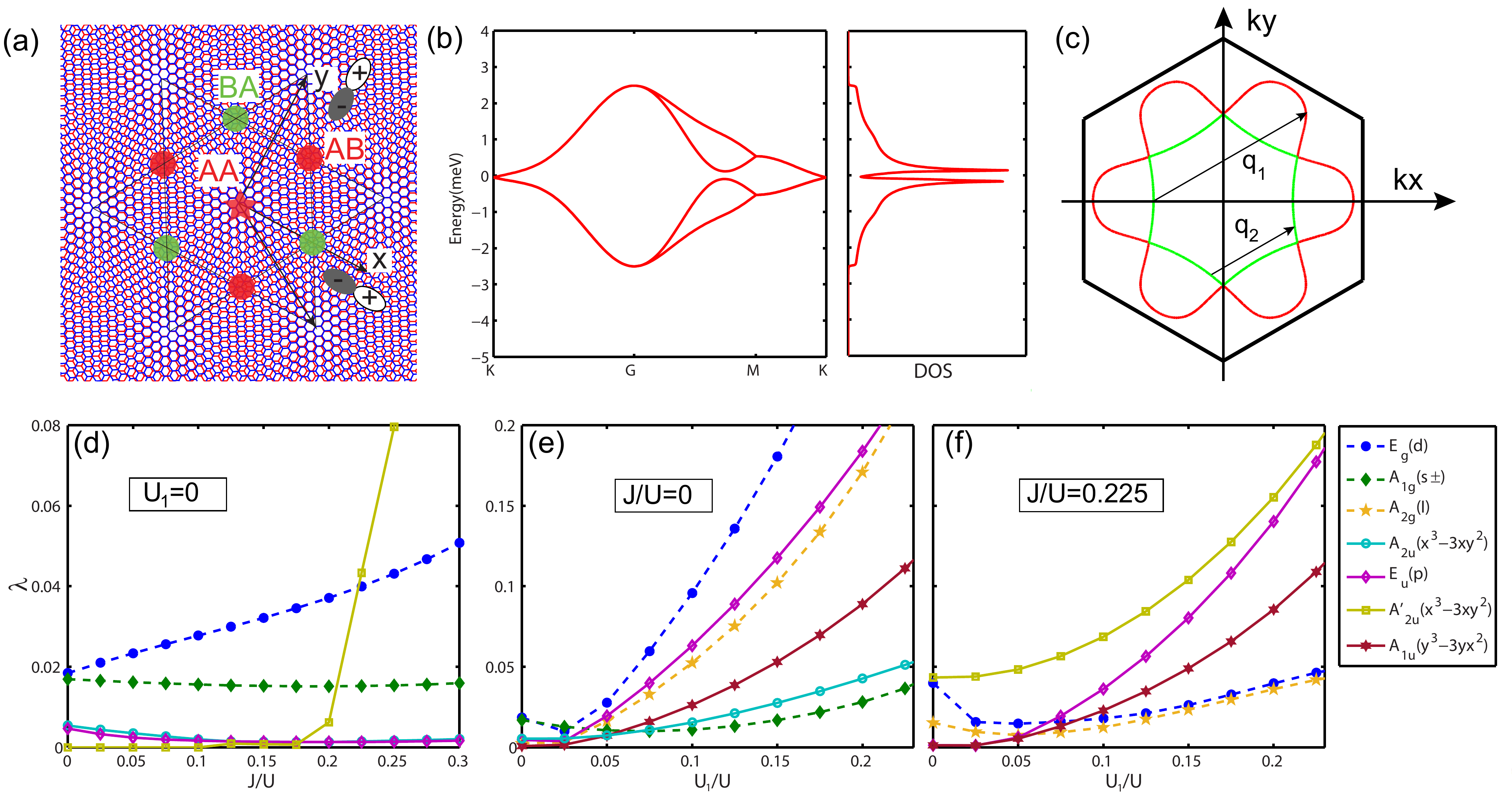}}
\caption{(color online) Atomic structure, band structure and pairing strength eigenvalues as a function of interaction for twisted bilayer graphene : (a) atomic structure for TBG. (b) band structure and density of states from effective tight binding model. (c) Fermi surfaces with 0.3 hole doping relative to half filling of valence bands ($n=1.7$). Pairing strength eigenvalues for the leading states: (d) $U=1.5$ meV and $U_{1,2,3}=0$, (e) $U=1.5$ meV, $J/U=0$ and $U_2=U_3=\frac{1}{2}U_1$ and (f) $U=1.5$ meV, $J/U=0.225$ and $U_2=U_3=\frac{1}{2}U_1$. In the irreducible representation, $g$ ($u$) denotes the even (odd) pairing state under inversion and is given in dashed (solid) lines. \label{NTBocc17}  }
\end{figure*}

 On the other hand, the tight-binding descriptions of many authors have aimed at a low-energy model of the flat bands alone, i.e., describing just the two bands (or four if the usual valley degree of freedom is taken into consideration) in terms of localized orbitals or Wannier functions\cite{Yuan2018PRB,Kang2018PRX,Koshino2018PRX}.
 Nevertheless, recent work has shown interesting additional topological aspects, in that the expected symmetry and topology of these flat bands generate obstructions to the use of exponentially localized WFs just for these bands\cite{Po2018PRX,Zou2018PRB,Song2018,Po2018-2}. One way out is to regard some of the symmetries as emergent ones at low energy, rather than the symmetries of TBG \cite{Hejazi2019}. The Wannier functions derived by reducing the symmetry then form a honeycomb lattice and have a propeller-like form, created by the superposition of three charge pockets at neighboring AA lattice spots (Fig.\ref{NTBocc17}(a))\cite{Koshino2018PRX,Po2018PRX,Kang2018PRX}. The extended shape of the Wannier functions then induces long-range hopping and interaction terms (even in the continuum model\cite{Pizarro2019}).

With superconductivity being a "low-energy" phenomenon (of course only when the "high-energy" electronic degrees are integrated out), we concentrate on here the nearly flat bands only. However, in contrast to the earlier work on electronic correlation-driven SC in TBG, we have to give up the on-site Hubbard-type of approach and consistently with the above discussion, have to include non-local, longer-range interactions. Key points of our results are that the non-local interactions not only increase the weight of higher lattice harmonics, but also have characteristic implications on the orbital structure of the pairing states. This should help in further sorting out the proposed different pairing mechanisms in TBG.

Based on a symmetry analysis, Yuan {\em et. al}\cite{Yuan2018PRB} proposed a  two-orbital TB model on the honeycomb lattice to describe the low-energy nearly flat bands of TBG, with two sublattices located at AB and BA stacking points, respectively, as shown in Fig.\ref{NTBocc17}(a). The proposed TB model reads as,
\begin{widetext}
\begin{equation}
H_0=\sum_{n=1}^{5}\sum_{\alpha\beta\nu}\sum_{\langle ij\rangle\in\{n\}}t_nc^\dag_{i\alpha\nu}c_{j\beta\nu}+\sum_{\alpha\nu}\sum_{\langle ij\rangle\in\{5\}}t^{\nu}_5\rho_{ij}c^\dag_{i\alpha\nu}c_{j\alpha\bar{\nu}}-\mu\sum_{i\alpha\nu}c^\dag_{i\alpha\nu}c_{i\alpha\nu},
\label{eqH}
\end{equation}
\end{widetext}
where $\alpha/\beta=A,B$ denotes the sublattice index, $\mu$ the chemical potential, the $n$ index stands for the $n$-th nearest neighbor (NN) bond, and $\nu$ is the $p_x$ and $p_y$ orbital degree. Here we consider the hopping up to the 5-th NN. $c^\dag_{i\alpha\nu}$ creates an electron in orbital $\nu$ on sublattice  $\alpha$ sublattice with unit cell index $i$. $\rho_{ij}=\pm1$ is a prefactor for the 5-th NN, and $t^x_5=-t^y_5$ (see Appendix \ref{modelkspace}).

Defining the spinor $\psi^\dag_{\bm{k}\sigma}=(c^{\dag}_{\bm{k}Ax\sigma},c^{\dag}_{\bm{k}Ay\sigma},c^{\dag}_{\bm{k}Bx\sigma},c^{\dag}_{\bm{k}By\sigma})$, the tight binding Hamiltonian takes the form
$H_{0}=\sum_{\bm{k}\sigma}\psi^\dag_{\bm{k}\sigma}h(\bm{k})\psi_{\bm{k}\sigma}$. The general Hamiltonian matrix in momentum space is delegated to the Appendix~\ref{modelkspace}. The Hamiltonian~\eqref{eqH} becomes block diagonal in the basis of $c^{\dag}_{j\alpha\xi}=c^{\dag}_{j\alpha x}+i\xi c^{\dag}_{j\alpha y}$ for the valley index $\xi=\pm$, highlighting the absence of intervalley coupling in the continuum model. The second term in Eq.~\ref{eqH} breaks inversion symmetry, yielding $D_3$ point group symmetry. A pseudo inversion symmetry, which flips both sublattices and orbitals, however, is still preserved~\cite{Tang2018} and, thus, the superconducting pairing states still decouple into spin singlet and spin triplet channels.

 Fitting the tight-binding model with up to the 5-th NN hopping to the band structure obtained from the continuum model with twisted angle $\theta=1.05^\circ$\cite{Koshino2018PRX}, the resulting band structure is given in Fig.\ref{NTBocc17}(b). Reasonable agreement is reached between continuum and tight-binding model (see Appendix \ref{modelkspace}). To achieve a perfect fitting of the continuum model, hopping parameters with distances up to 9$L_M$ should be included\cite{Koshino2018PRX}, where $L_M$ is the lattice constant of the Moir\'e unitcell. Despite narrower bandwidth and quantitative differences around the $\Gamma$ point, this model captures the main features of band structures from the continuum model.

Furthermore, the obtained double-peak density of states and triangular Fermi surfaces are consistent with those from the continuum model. This model will be adopted in the following calculations. The Fermi surface near half filling of valence bands (0.3 hole doped) is displayed in Fig.\ref{NTBocc17}(c). The Fermi surfaces from the two valleys are trigonal and related by $C_{2y}$ rotation as well as time reversal.


\begin{figure*}[tb]
\centerline{\includegraphics[width=2.0\columnwidth]{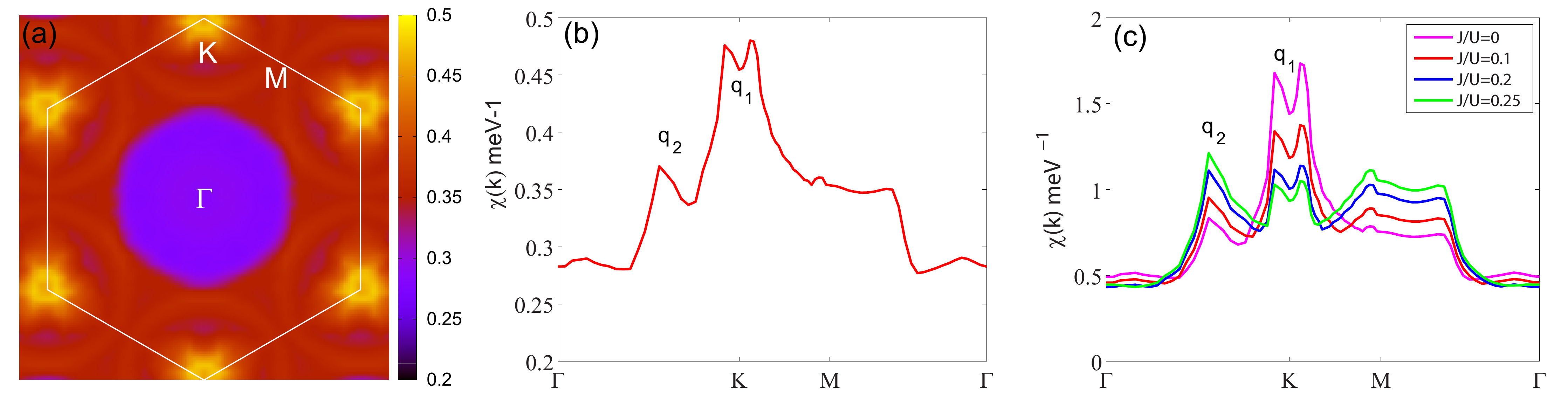}}
\caption{(color online) Distribution of the largest eigenvalues for bare susceptibility matrices $\chi_0(\bm{k})$ at $n=1.7$. (a) in Brillouin zone (b) along high-symmetry path. (c) RPA susceptibility along high-symmetry path with $U=$1.5 meV, plotted for increasing $J/U$ ratios.   \label{susbzNTBocc17}  }
\end{figure*}

\begin{figure}[tb]
\centerline{\includegraphics[width=1\columnwidth]{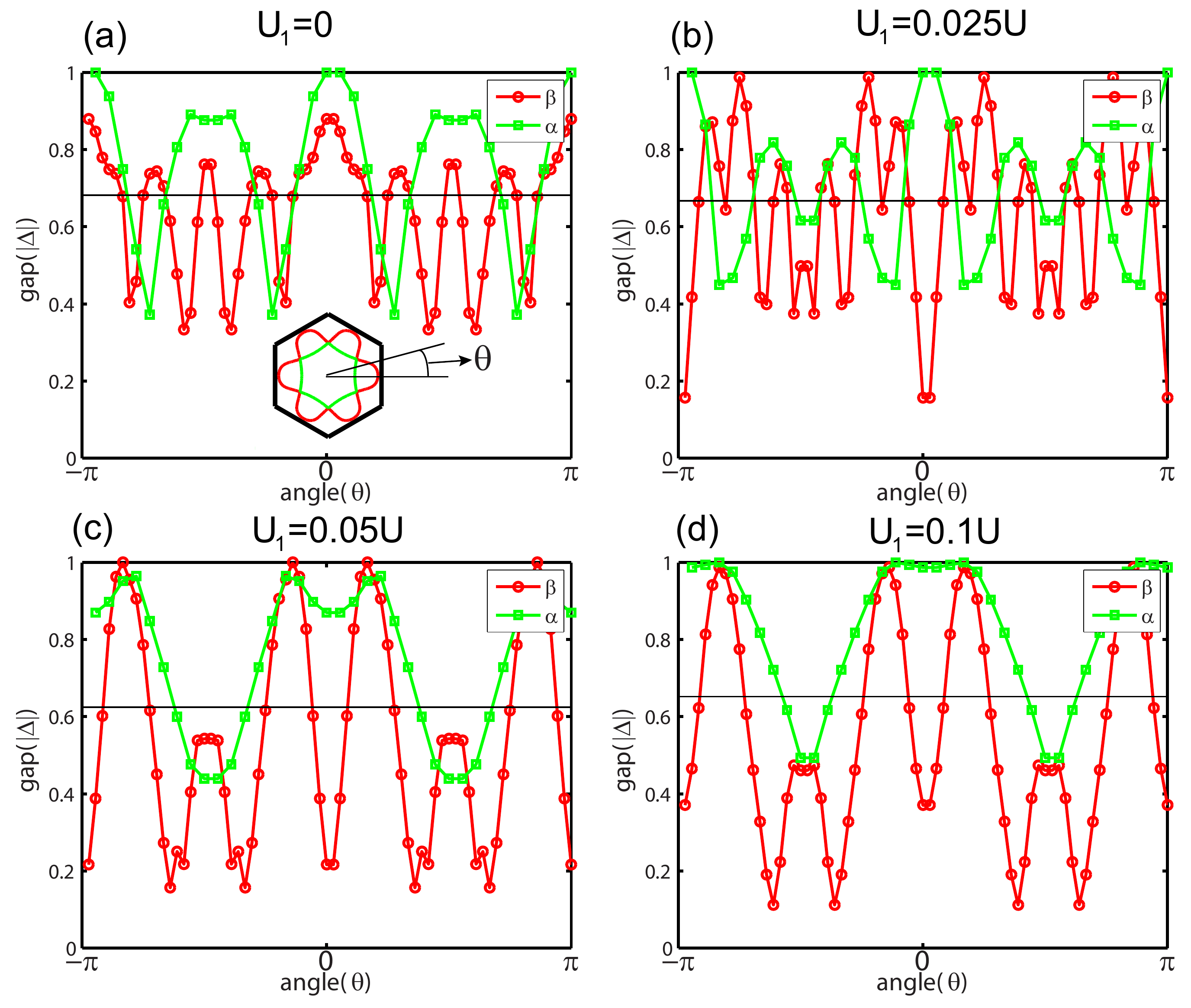}}
\caption{(color online) The $d+id$ gap functions on Fermi surfaces as a function of $U_1$. The ratio of nonlocal interactions is $U_1:U_2:U_3=2:1:1$, $U=1.5$ meV and $J/U=0$. The black solid line denotes the average gap. The gap anisotropy is given by the variance of gap function divided by the mean and the obtained gap anisotropies are 0.26, 0.31,0.42 and 0.43 for (a)-(d), respectively. The gap anisotropies for the corresponding $s_{\pm}$ states with the same interactions are 0.72, 0.74, 0.76 and 0.80. The gap anisotropy is defined in Ref.\onlinecite{Platt2013}.  \label{gapdidV} }
\end{figure}


\section{pairing symmetries from onsite and long-range interactions} \label{pairingRPA}
 The fact that the Wannier orbitals are rather extended in Model~\eqref{eqH}, inescapably implies that the nonlocal interactions should be included\cite{Koshino2018PRX}. We consider nonlocal Hubbard type density-density interactions up to third-nearest neighbors, and full onsite intra- and inter-orbital terms, Hund's coupling and pairing hopping type interactions, i.e.
\begin{widetext}
\begin{eqnarray}
H_{int}&=&U\sum_{i\mu}n_{i\mu\uparrow}n_{i\mu\downarrow}+U'\sum_{i,\mu<\nu}n_{i\mu}n_{i\nu}+J\sum_{i,\mu<\nu,\sigma\sigma'}c^{\dag}_{i\mu\sigma}c^{\dag}_{i\nu\sigma'}c_{i\mu\sigma'}c_{i\nu\sigma}
+J'\sum_{i,\mu\neq\nu}c^{\dag}_{i\mu\uparrow}c^{\dag}_{i\mu\downarrow}c_{i\nu\downarrow}c_{i\nu\uparrow}\nonumber\\
&&+\sum_{n=1}^{3}\sum_{\alpha\beta, \nu}\sum_{\langle ij\rangle\in\{n\}}U_nn_{i\alpha\nu} n_{j\beta\nu}+\sum_{n=1}^{3}\sum_{\alpha\beta, \nu<\mu}\sum_{\langle ij\rangle\in\{n\}}U'_n n_{i\alpha\nu}n_{j\beta\mu},
\label{interaction2}
\end{eqnarray}
\end{widetext}
where $n_{i\alpha}=n_{\alpha\uparrow}+n_{\alpha\downarrow}$. $U$, $U'$, $J$ and $J'$ represent the onsite intra-orbital, inter-orbital repulsion, Hund's coupling and pair-hopping terms, respectively. $U_n$ and $U'_n$ are the $n$-th NN intra- and inter-orbital repulsion, the decay rate of which with distance is dictated by  the three-peak structure of the Wannier functions around the AA site. As the two orbitals belong to one two-dimensional irreducible representation, interactions still preserve crystal symmetries by adopting the Kanamori relations $U=U'+2J$ and $J=J'$. Our motivation to include a Hund's coupling term is as follows: it has been convincingly argued, corroborated by dynamical mean-field results for the metal-insulator transition in the two-band Hubbard model\cite{Pruschke2005}, that in multi-orbital compounds the renormalization of the kinetic energy and the incoherence of the normal state is tied up with $J$\cite{Haule2009}. Competing magnetic states are also determined by Hund's rule coupling.

Near half filling of valence bands (0.3 hole doped), as shown in Fig.\ref{susbzNTBocc17}(a) and (b), the bare susceptibility shows a predominant peak around the K point, which is mainly attributed to the interpocket nesting $\bm{q}_1$, depicted in Fig.\ref{NTBocc17}(c). In addition, the intrapocket nesting $\bm{q}_2$ will contribute a peak near the midpoint of $\Gamma K$. This peak lying closer to the $\Gamma$ point, indicates ferromagnetic fluctuations. It will get enhanced with including Hund's rule coupling as discussed below. Both peaks in the spin susceptibility are of central relevance when interactions are considered within RPA. For interaction greater than a critical value $U_c$, the spin susceptibility $\chi(\bm{q}_{1/2})$ will diverge and indicate a spin density wave (SDW) instability. Below $U_c$, superconductivity can emerge triggered by spin fluctuations.

Near $T_c$, the gap function can be obtained by solving the linearized gap equation,
\begin{eqnarray}
-\sum_{j} \oint_{C_j} \frac{dk'_{\|}}{4\pi^2v_F(\textbf{k}')} \Gamma_{ij}(\textbf{k},\textbf{k}') \Delta_{\alpha}(\textbf{k}')=\lambda_{\alpha}\Delta_{\alpha}(\textbf{k}), \label{vertex}
\end{eqnarray}
where $\lambda_{\alpha}$ denotes the pairing strength for the gap function $\Delta_{\alpha}(\textbf{k})$, $i,j$ are the Fermi surface indexes, and the effective interaction vertex $\Gamma_{ij}$ (Appendix \ref{rpamethod}) is the symmetric (antisymmetric) part of the full interaction in the singlet (triplet) channel. Expression~(\ref{vertex}) is not specific to RPA, but emerges in a similar form for perturbative RG via Kohn Luttinger\cite{Kohn1965}, functional renormalization group\cite{Metzner2012,Platt2013}, and other approaches. Due to the virtue of being an eigenvalue equation, every symmetry of the effective interaction $\Gamma$ implies a transformation of $\Delta_{\alpha}(\textbf{k})$ as an irreducible representation of the same symmetry. \emph{The orbital- and sublattice-specific Fourier decomposition of $\Delta_\alpha(\textbf{k})$ is what we define to be the HFP of superconducting pairing}. Due to the presence of pseudo inversion symmetry without intervalley coupling, the pairing states are classified according to the irreducible representations of $D_{3d}$.

 Consider first onsite interactions only, and fix $U=1.5 meV$, while varying $J/U$. The RPA results for the leading  pairing strength eigenvalues $\lambda_{\alpha}$ are displayed in Fig.\ref{NTBocc17}(d). With $J/U<0.2$, we find a distinct separation of
the two-fold degenerate leading eigenvalues from $E_{g}$ ($d$-wave) and a non-degenerate eigenvalue from $A_{1g}$ ($s_{\pm}$-wave) from the other subleading ones.
They are rather close for small $J/U$; for increasing $J/U$, the pairing eigenvalue of the $d$-wave state increases while that of the $s_{\pm}$ state stays nearly constant. The gap functions of the before-mentioned three pairing states are given in the Appendix~\ref{gapfunction}. For the $A_{1g}$ state, the related superconducting gap is nodeless but anisotropic, due to higher harmonics in the HFP, already present for onsite interactions only, and it features a sign change between the inner pocket and outer pocket\cite{LiuZ2018}.

We decompose the gap functions of the two-fold degenerate $E_g$ state into $d_{x^2-y^2}$ and $d_{xy}$ symmetries which reach a maximum of amplitude near the lobes. In order to maximize the condensation energy and the gap function, a mean field analysis of the doubly degenerate $E_g$ pairing yields the formation of a nodeless $d+id$ state which spontaneously breaks time reversal symmetry. Both the $d+id$ state and the $A_{1g}$ state, are nodeless and of unconventional origin. They satisfy the condition that the superconducting gaps connected by the nesting vector $\bm{q}_1$ have a sign change in the singlet channel.

For $J/U>0.2$, the eigenvalue of an $A'_{2u}$ state in the triplet channel increases rather abruptly, and becomes dominant. The corresponding gap function of this $A'_{2u}$ triplet state, dominant at large $J/U$, is isotropic and nodeless. It has opposite signs for Fermi surfaces originating from two valleys\cite{Tang2018}. Any intervalley scattering, however, can couple these two Fermi surfaces, and will naturally introduce nodes along the $\Gamma$-K line. This pairing state is supported by intra-pocket nesting.

 It is instructive to reveal the mechanism for this abruptly enhanced triplet pairing qualitatively using diagrams (Appendix \ref{bubbleonsite} ). The first two diagrams give the repulsive interaction and the latter bubble diagrams the attractive interaction. When Hund's rule coupling is small, the first order diagrams dominate and the effective interaction is repulsive. Therefore, orbital singlet pairing will not be favored. With increasing $J/U$, $\chi(\bm{q}_2)$ gets enhanced significantly, as shown in Fig.\ref{susbzNTBocc17}(c). Simultaneously, the first-order repulsive interaction decreases, but the contribution from latter bubble diagrams increases rapidly. When $J/U>0.2$, the contribution from the bubble diagrams can overcome the first-order term, which gives rise to the orbital-singlet spin-triplet onsite pairing. The superconducting gaps connected by the $\bm{q}_2$ vector should have the same sign due to the effective attractive interaction.

In the next step, we approach a more realistic scenario for TBG by investigating the effect on the pairing states due the nonlocal interactions, and fix the ratio to $U_1:U_2:U_3=2:1:1$ and $U_n=U'_n$ motivated by the three-peak structure at AA site for each Wannier orbital\cite{Koshino2018PRX}. For the case where Hund's rule coupling and pairing hopping are negligible ($J/U=0$), the pairing strength eigenvalues $\lambda_{\alpha}$ as a function of $U_1/U$ are shown in Fig.\ref{NTBocc17}(e). Note that, while the pairing strength is plotted in Fig.\ref{NTBocc17} (e) and (f) as a function of $U_1/U$, this includes automatically also long-range interactions up to $U_3$. With increasing nonlocal interactions in the singlet channel, the $d$-wave state is first slightly suppressed and then gets enhanced rapidly while in the triplet channel, $E_{u}$($p$-wave ) and $A_{1u}$ ($f_{y^3-3yx^2}$-wave) states also get enhanced. Although the dominant pairing state is always $E_g$, we find that the corresponding gap functions exhibit a significant change with increasing $U_1$, as shown in Appendix \ref{gapV}. So does the gap function for the leading pairing state in the spin-triplet channel. These findings suggest that the IR of a pairing state is not enough to characterize its precise shape and the detailed structure of the HFP need to be further taken into consideration.

For a typical Hund's rule coupling $J/U=0.225$ in Fig.\ref{NTBocc17}(f), we find that singlet pairing states are suppressed, while triplet states exhibit a substantial enhancement with increasing nonlocal interactions. The triplet gap functions of the leading states display pronounced variations with increasing interactions (Appendix~\ref{gapV}), which suggests a significant change of the real-space pairing structure, and hence their HFP.

Summarizing this section, we find that, with onsite interactions, $d$-wave and $s_{\pm}$ pairing states are dominant with small Hund's rule coupling ($J/U<0.2$). On the other hand, when $J/U>0.2$, an $f$-wave pairing state is favored. These results are consistent with previous studies\cite{Liu2108PRL,Tang2018,LiuZ2018}. With further including nonlocal interactions, $E_g$ i.e. $d+id$ pairing is the dominant channel for small $J/U$, but the corresponding gap functions exhibit a significant variation with increasing $U_1$. With a larger $J/U$, where a value around 0.2 seems not unreasonable for TBG, the $f$-wave pairing state dominates and the corresponding gap anisotropy is strongly enhanced with increasing $U_1$. These findings support our notion, that the nonlocal interactions have a significant effect on HPF. In the following section, we will trace the HPF evolution of the dominant pairing states to the variation of nonlocal interactions in more detail.

\begin{table}[t]
\caption{\label{onsite} Allowed onsite pairing states on two-orbital honeycomb lattice. "+"("-") in orbital and spin space represents triplet (singlet).  }
\renewcommand{\multirowsetup}{\centering}
\begin{tabular}{ccccc}
\hline
\hline
IR(orbital) & orbital & spin & IR(band) & Matrix($F(\bm{k})$)\\
   \hline
 $A_1$  & + & - & $A_{1g}(s)$ & $is_{2}\sigma_{0,3}\tau_0 $   \\
 $A'_1$ & - & + & $A'_{2u}$($f_{x^3-3xy^2}$)& $s_{1}\sigma_{0,3}i\tau_2 $ \\
 $E_1 $ & + & - & $E_{g1}(d_{x^2-y^2})$ & $is_{2}\sigma_{0,3}\tau_3$       \\
 $E_2 $ & + & -  & $E_{g2}(d_{xy})$ & $is_{2}\sigma_{0,3}\tau_1 $  \\
\hline
 \hline
\end{tabular}

\end{table}

\section{ Pairing analysis in real space}  \label{realspacepairing}
Aside from a mere harmonic Fourier analysis of the pairing wave function, the HFP of a pairing state in general also involves sublattice, orbital, and spin degrees of freedom. Introducing the spinor $\Psi^\dag_{\bm{k}}=(\psi^\dag_{\bm{k}\uparrow},\psi^\dag_{\bm{k}\downarrow})$, we write down the full classification of pairing states according to point group symmetry. The pairing function reads
\begin{eqnarray}
\hat{\Delta}_\alpha=\sum_{\bm{k}ijln}f^n_{ijl}(\bm{k})\Psi^\dag_{\bm{k}}s_{i}\otimes \sigma_j \otimes \tau_l [\Psi^\dag_{-\bm{k}}]^T,
\end{eqnarray}
where $\bm{s}$, $\bm{\sigma}$, $\bm{\tau}$ denote Pauli matrices in the spin, sublattice and orbital space, $f^n_{ijl}(\bm{k})$ is the $n$-th NN lattice harmornic and $i,j,l=0,1,2,3$. In the above basis, the representation matrices of two generators of $D_3$ are: $D(C_{2y})=is_{2}\sigma_1\tau_3$ and $D(C_{3z})=e^{-i\frac{\pi}{3}s_3} \sigma_1  e^{\frac{2i\pi}{3}\tau_2}$. We rewrite the pairing state as $\hat{\Delta}_\alpha=\sum_{\bm{k}}\Psi^\dag_{\bm{k}}F(\bm{k}) [\Psi^\dag_{-\bm{k}}]^T$, which under a point group operation $g$ transforms as,
\begin{eqnarray}
\hat{P}_g\hat{\Delta}_{\alpha}\hat{P}^{-1}_g&=&\sum_{\beta}\hat{\Delta}_{\beta}O_{\beta\alpha}(g)\nonumber\\
&=&\sum_{\bm{k}}\Psi^\dag_{\bm{k}}D(g)F(g^{-1}\bm{k}) D^T(g)[\Psi^\dag_{-\bm{k}}]^T,
\end{eqnarray}
where $D(g)$ and $O(g)$ are the representation matrices for Bloch and pairing states. For a one-dimensional IR, we have $\hat{P}_g\Delta_{\alpha}\hat{P}^{-1}_g=\eta_{g,\alpha}\Delta_\alpha$, where the eigenvalue can be obtained by solving $D(g)F(g^{-1}\bm{k}) D^T(g)=\eta_{g,\alpha}F(\bm{k})$. In the following, we first classify the pairing in orbital and sublattice space and then combine them with lattice harmonics.

Due to spin rotation invariance for TBG, we only need to consider the spin-singlet and spin-triplet states both present in the $S_z=0$ pairing sector. First, we discuss the pairing states in orbital space within the same sublattice. Due to the fermionic antisymmetry, only eight onsite pairing states are allowed and listed in Table \ref{onsite}, where $\sigma_0$ ($\sigma_3$) represent the same (opposite) gap functions for the two sublattices. Among them, $A_1$ pairing is spin-singlet orbital-triplet and $A'_1$ is spin-triplet orbital-singlet. They correspond to $s$-wave and $f_{x^3-3xy^2}$-wave pairing in band space. The latter two pairings $E_{1,2}$ are spin-singlet, orbital-triplet states, and relate to $d$-wave pairing in band space, belonging to a two-dimensional IR. Assuming the absence of inter-valley scattering and time-reversal symmetry, only the first two entries of Table \ref{onsite} can be present in TBG. It is straight-forward to extend this classification to bond-pairing by multiplying lattice harmonics with the same sublattice.

 We further consider the intersublattice pairing and the lattice harmonic $f_{inter}(\bm{k})$. Pairing in sublattice space can be written as $f_{inter}(\bm{k})\sigma_+\pm f_{inter}(-\bm{k})\sigma_-$, corresponding to the sublattice triplet or singlet states. Combing it with oribital and spin pairings, intersublattice pairing can be obtained and classified (the details can be found in Appendix \ref{fitting}). We provide the pairing states for NN, NNN and TNN bonds and the corresponding IRs in band space in the Table \ref{pairrealspace} of Appendix \ref{fitting}. One significant feature is that the IR in real space and band space can generally differ. For example, $A'_1$ onsite pairing corresponds to an $f$-wave ($A'_{2u}$) pairing in band space.

In order to obtain the HFP from the pairing states obtained through RPA, we further project the real-space pairing states onto the Fermi surface and treat them as the bases for the decomposition of RPA gap functions. The given IR then specifies the basis set to use, where we only keep the first four basis states. For $J/U=0$ and $U_1=0$, the relevant pairing states ($s_{\pm}$, $d$ and $p$-wave states) are mainly attributed to NN and NNN pairing (Appendix \ref{realpairing}). With increasing nonlocal interaction, not only the Cooper pairing distance increases and the HFP weights are shifted to higher lattice harmonics, but also the orbital part of pairing changes (except for the $s_{\pm}$ state).

For the dominant $d$-wave state at $U_1$ exceeding $0.1U$, orbital singlet pairing on the NNN bond with lattice harmonics according to $E'_1$ and $E'_2$ (odd in $\bm{k}$ space) provide the dominant contribution to the HFP, in sharp contrast to the intra-orbital pairing with even lattice harmonics. As the two-fold degenerate states tend to form a $d+id$ state to gain condensation energy, we plot the gap functions of the $d+id$ state for four typical values of $U_1$ in Fig.\ref{gapdidV}. With increasing $U_1$, the gap functions become more anisotropic, and the depth of the gap minima gets enhanced. The subdominant $A_2$ ($I$-wave) state is mainly attributed to orbital-singlet pairing on NNN bond with an $A'_1$ lattice harmonic. Therefore, in the spin-singlet channel, the orbital-singlet pairing with an odd harmonic is favored.

In the spin-triplet channel, with large nonlocal interactions, the $p$-wave pairing state is mainly attributed to the intra-orbital pairing on NNN bonds, composed by $E'_2$ harmonics. For $J/U=0.225$, spin-triplet pairing dominates over spin-singlet pairing, as shown in Fig.\ref{NTBocc17}(f). The dominant $A'_{2u}$ ($f$-wave) state possesses a considerable onsite orbital-singlet pairing, and the corresponding NN and third NN (TNN) pairing components of the HFP also increase with increasing $U_1$, \emph{which leads a more anisotropic gap function}. For $J/U=0$, nonlocal interactions suppress intraorbital (orbital-triplet) pairing but promote interorbital (orbital-singlet) pairing on the NNN bond for a $d$-wave spin-singlet state. It is just the opposite for $p$-wave spin-triplet state, where intra-orbital pairing on NNN bond dominates.

\begin{figure}[tb]
\centerline{\includegraphics[width=1\columnwidth]{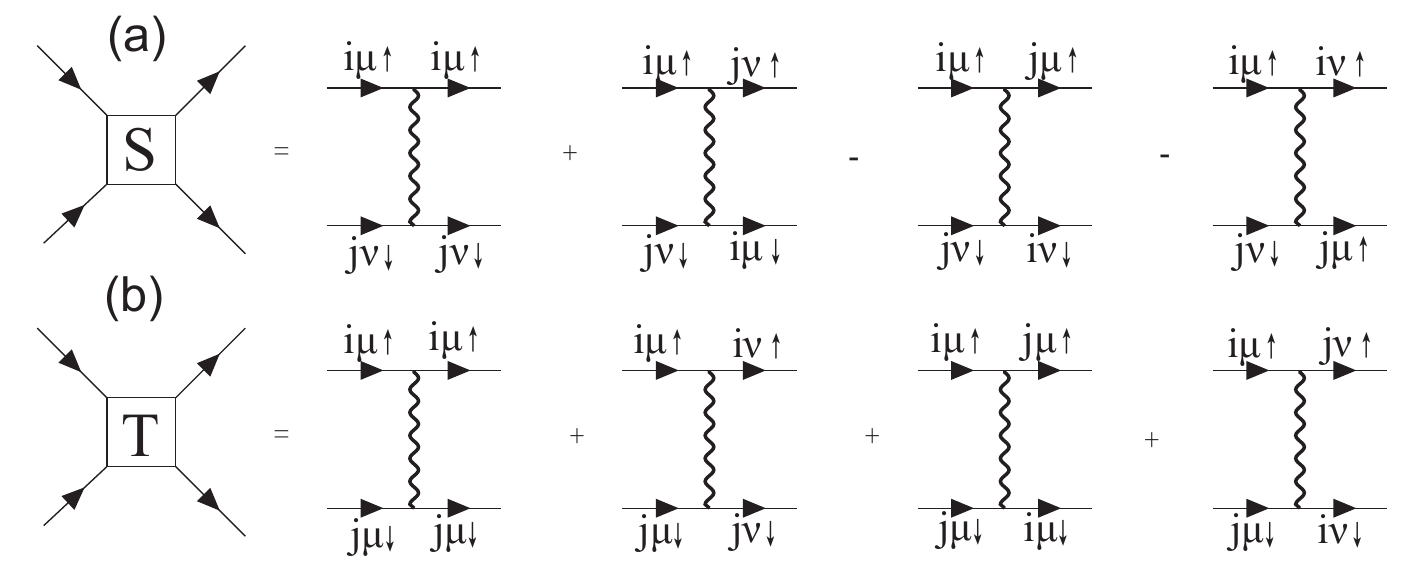}}
\caption{(color online) Effective pairing interaction for orbital singlet (a) and triplet (b) in spin singlet channel on NNN bond. $i,j$ are the lattice site and $\mu,\nu$ label different orbitals. \label{multiorbitaleff}  }
\end{figure}

In the following, we provide an qualitative explanation from the Feynman diagram perspective. In the spin-singlet channel, the effective pairing interactions are even(odd) in space for orbital-triplet(singlet) channels, as shown in Fig.\ref{multiorbitaleff}. The first order diagrams contribute repulsive interactions. However, in the second order, the bubble diagram can be attractive (first term in Fig.\ref{multiorbitaleff}(a) and (b) ). Except the first terms in Fig.\ref{multiorbitaleff} (a) and (b), the other terms can only have ladder diagrams from the onsite and nonlocal interactions in the second order, hence are repulsive. Due to the intrinsic symmetry properties of the effective interactions, the last two terms in Fig.\ref{multiorbitaleff}(a) and (b) have the opposite signs. Therefore, including nonlocal interactions will relatively suppress orbital-triplet pairing but enhance orbital-singlet pairing. This explains the obtained HFP change in our calculations. Similarly, we can explain the dominant intra-orbital pairing for the $p$-wave state in spin-triplet channel. With a large Hund's rule coupling, $\chi(\bm{q}_2)$ dominates in Fig.\ref{susbzNTBocc17}(c) and onsite orbital-single pairing is the leading state.

In this section, starting from the effective pairing vertex we studied how the pairing wavefunction unfolds in its sublattice, orbital and spin degrees-resolved Fourier profiles, i.e. the harmonic fingerprint. We first classify the real-space pairing in the two-orbital honeycomb lattice and then find that nonlocal interactions strongly affect both the orbital structure and the higher harmonics of the pairing state. In the final section, an STM methodology will be discussed which, in our opinion, is ideally suited for revealing these detailed microscopic insights into the pairing states.

\section{discussion and conclusion}\label{conclusion}

 Let us first comment on the electronic structure and the need for including longer-range interactions. In the effective model for TBG, standard lattice symmetries have to be obeyed and the observed real-space charge density pattern, peaking at the Moir\'e triangle sites, should be recovered. The nearly flat bands with a bandwidth of order $\sim$ 10 meV are well separated by a gap $\Delta E$ experimentally of order 10-20 meV from the higher bands. Then standard theory on the construction of Wannier functions (WFs) by W. Kohn and others \cite{Kohn1973,Cloizeaux1964} would tell us, that we can construct exponentially localized WFs with an exponential decay constant of order
 $r\approx\Delta E^{-1/2}$.  Because of the small gap $\Delta E$, these WFs are necessarily rather spread out. However, it is also known from earlier work\cite{Cloizeaux1964} that, if one includes higher bands in a "band complex"  now separated by a significantly larger gap $\Delta E_2$ from the rest of even higher bands, then the decay constant is of order $r\approx\Delta E_2^{-1/2}$. In this case, the WFs can become obviously much more localized.

    To overcome the "fragile topology" of the nearly flat bands, i.e. the obstruction to construct exponentially localized WFs, recent work suggested adding a particular set of bands, a kind of "band complex" to the flat-band model. As mentioned already in the Introduction, this approach is extremely complicated from a numerical point of view, based on the extremely large Moir\'e supercell with about 12000 atoms. Even for such "a faithful TB model" in TBG, the WFs can still be extended if $\Delta E_2$ is relatively small and, therefore, the nonlocal interactions can also be significant and important. Our studies may then still be taken to point out crucial effects on the HFP for the pairing states with nonlocal interactions.

Let us then discuss our results for SC pairing and possible competing Fermi-surface instabilities in TBG, also commenting on some other work. There is a general aspect concerning the HFP of pairing and a more specific one relating to TBG. If one considers as the simplest 2D case the square-lattice Hubbard model with NN hopping and onsite interaction, then e.g. numerical Quantum Monte Carlo calculations demonstrate that holes on NN (next NN) sites have an attractive (repulsive) interaction\cite{Scalapino1995}. Therefore, if one further includes NN Coulomb repulsion, depending on its strength, SC may be completely suppressed and replaced by a charge-density-wave order. Even if SC prevails, the dominant pairing interaction will be shifted to next NN sites and in general, longer-ranged Coulomb interactions will induce longer-ranged Cooper pairing tendencies. When this picture is supplemented with the orbital, sublattice and spin decomposition of the SC pairing state (as dictated by the general Hamiltonian in Eq.\ref{interaction2}), we arrive at the notion of a harmonic fingerprint of SC order and its physical relevance. Of course, in realistic material situations the unavoidable presence of longer-ranged Coulomb interactions further emphasizes the usefulness of such a point of view.

Earlier work on the possibility of SC in TBG just considers the SU(4) symmetric Hubbard interaction\cite{XuCK2018PRL}. It found a chiral spin triplet $d+id$ topological SC state bordering the correlated insulating state near half-filling. In the next step of sophistication, the interactions between the electrons are still assumed local (as the first line in our Eq.\ref{interaction2}), but contain the on-site intra- and inter-orbital contribution, Hund's rule and pair-hopping terms. Local Hunds's rule coupling favors a pairing between two electrons in an orbital singlet. By Fermion asymmetry, the spin part must be a triplet and the pairing function projected in the band basis displays $f$-wave symmetry (see our Fig.\ref{susbzNTBocc17} and the discussion in Sec. III). Summarizing and consistent with previous studies\cite{Liu2108PRL,Tang2018,LiuZ2018}, including only these local interactions, we find  $d+id$-wave and $s_{\pm}$-wave pairing for a small Hund's rule coupling ($J/U<0.2$). For a large Hund's rule coupling ($J/U>0.2$), it is replaced by an $f$-wave pairing state. For this, we provide also a diagrammatic explanation in Appendix \ref{bubbleonsite}. Our crucial point is that the inclusion of unavoidably longer-ranged interactions in the starting Hamiltonian, has a significant and characteristic effect on the HFP, which should, in principle, be amenable to experimental studies. With inclusion of the nonlocal interactions, although the $d+id$-wave and $f$-wave pairing still dominate for small and large Hund's rule coupling, the gap anisotropy  gets significantly enhanced with increasing nonlocal interactions, originating from their effect on both orbital structure and the higher harmonics of the pairing state. Therefore, longer-ranged interactions have a significant impact on the real-space pairing (as discussed already for the Hubbard model on a square lattice) and hence their HFP. Our studies may be taken to point out crucial effects on the HFP.

This harmonic fingerprint should be reflected in the experimentally observable gap function
from an elementary spectroscopic analysis only sensitive to the quasiparticle density of states. In particular, for the powerful scanning tunneling microscopy (STM) technique, recent work has shown that the obtained conductance spectra, i.e. $dI/dV$ spectra can be used to distinguish $f$-wave and $d+id$ pairing states for a hexagonal lattice\cite{Elster2015}.
For the $f$-pairing nodal state the conductance curves contain zero-energy peaks due to conserved time-reversal symmetry (TRS). This is in sharp contrast to a possible $d+id$ pairing state, here, one expects a gap and a kind of sunflower structure in the $dI/dV$ characteristics for the topological pairing $d+id$ phase with broken TRS(zero differential conductance until the minimal pairing potential is reached)\cite{Elster2015}. In contrast to this, the differential conductance increases continuously with energy from zero for the gapless $f$-pairing phases.

 Moreover, it has been demonstrated for both $d+id$ and $f$-wave SC gaps, that their conductance spectra characteristically change when higher harmonics enter the pairing function\cite{Elster2015}. This indeed implies that the HFP can be experimentally detected. Additionally, the numbers of bound states induced by impurities can also be used to distinguish $s$-, $d+id$- and $p+ip$-wave pairing states\cite{YangH2019}.

In conclusion, the HFP of a superconducting state establishes first of all a kind of microscopic fingerprint to analyse its univerlying pairing mechanism beyond mere point-group symmetry. As we have shown for a microscopic model of TBG,  nonlocal interactions have a significant effect on the HFP. The enhanced range of interactions tends to characteristically change the orbital structure and induces higher lattice harmonics in the pairing states, and, as such, more anisotropic gaps. Thus, the HFP provides, in principle, a powerful tool to investigate the microscopic origin of electronic pairing in TBG, and any potentially unconventional superconducting scenario, where multi-orbital effects and longer-range interactions conspire in building up the SC phase.

\acknowledgments
We thank A. V. Chubukov, M. Greiter, and N. Yuan for discussions. The work in W\"urzburg is funded by the Deutsche
Forschungsgemeinschaft (DFG, German Research Foundation) through Project-ID 258499086 - SFB 1170 and through the W\"urzburg-Dresden Cluster of Excellence on Complexity and Topology in Quantum Matter --\textit{ct.qmat} Project-ID 39085490 - EXC 2147.

\clearpage
\begin{widetext}
\appendix
\section{two-orbital tight binding model in momentum space for twisted bilayer graphene }\label{modelkspace}

Based on a point group argument ($D_3$), N. Yuan  \emph{et. al}\cite{Yuan2018PRB,Koshino2018PRX} proposed a four-band effective model to describe the low-energy electronoic structure of twisted bilayer graphene . The effective lattice is a honeycomb lattice with two sublattices at AB and BA spots and there are two degenerate $E$ orbitals with $p_{x,y}$ symmetries on each sublattice.
For a two-orbital model on this honeycomb lattice, a general tight binding model in momentum space can be written as,
\begin{eqnarray}
H_{TB}=\sum_{\alpha\beta}\sum_{\mu\nu\sigma}h^{\alpha\beta}_{\mu\nu}(\mathbf{k})c^\dag_{\alpha\mu\sigma}(\bm{k})c_{\beta\nu\sigma}(\bm{k}),
\end{eqnarray}
where $\alpha/\beta=A,B$ is the sublattice index and $\mu/\nu=1,2$ represent the $p_x$ and $p_y$ orbitals, respectively.
$c^\dag_{\alpha\mu\sigma}$ creates a spin $\sigma$ electron in the $\mu$ orbital on the $\alpha$
sublattice with momentum $\bm{k}$. The nonzero matrix elements $h^{\alpha\beta}_{\mu\nu}(\mathbf{k})$ for the NN and NNN hoppings are given by,
\begin{eqnarray}
h^{AA/BB}_{11}(\mathbf{k})&=&\epsilon_{p_x}+2T^{22}_{11}cos(k_y A_0)+(T^{22}_{11}+3T^{22}_{22})cos\frac{\sqrt{3}}{2}k_x A_0cos\frac{1}{2}k_y A_0,\\
h^{AA/BB}_{22}(\mathbf{k})&=&\epsilon_{p_y}+2T^{22}_{22}cos(k_y A_0)+(3T^{22}_{11}+T^{22}_{22})cos\frac{\sqrt{3}}{2}k_x A_0cos\frac{1}{2}k_y A_0,\\
h^{AA/BB}_{12/21}(\mathbf{k})&=&\sqrt{3}(T^{22}_{11}-T^{22}_{22})sin\frac{\sqrt{3}}{2}k_x A_0sin\frac{1}{2}k_y A_0,\\
h^{AB}_{11}(\mathbf{k})&=&t^{22}_{11}e^{\frac{ik_xA_0}{\sqrt{3}}}+\frac{1}{2}(t^{22}_{11}+3t^{22}_{22})e^{-\frac{ik_xA_0}{2\sqrt{3}}}cos(\frac{k_yA_0}{2}),\\
h^{AB}_{12/21}(\mathbf{k})&=& -\frac{\sqrt{3}i}{2}(t^{22}_{11}-t^{22}_{22})e^{-\frac{ik_xA_0}{2\sqrt{3}}}sin(\frac{k_y A_0}{2}),\\
h^{AB}_{22}(\mathbf{k})&=&t^{22}_{22}e^{\frac{ik_xA_0}{\sqrt{3}}}+\frac{1}{2}(3t^{22}_{11}+t^{22}_{22})e^{-\frac{ik_xA_0}{2\sqrt{3}}}cos(\frac{k_yA_0}{2}).
\end{eqnarray}
The third NN (TNN) hopping terms are,
\begin{eqnarray}
h^{AB}_{11}(\mathbf{k})&=&p^{22}_{11}e^{-\frac{2ik_xA_0}{\sqrt{3}}}+\frac{1}{2}(p^{22}_{11}+3p^{22}_{22})e^{\frac{ik_xA_0}{\sqrt{3}}}cos(k_yA_0),\\
h^{AB}_{12/21}(\mathbf{k})&=& \frac{\sqrt{3}i}{2}(p^{22}_{11}-p^{22}_{22})e^{\frac{ik_xA_0}{\sqrt{3}}}sin(k_y A_0),\\
h^{AB}_{22}(\mathbf{k})&=&p^{22}_{22}e^{\frac{-2ik_xA_0}{\sqrt{3}}}+\frac{1}{2}(3p^{22}_{11}+p^{22}_{22})e^{\frac{ik_xA_0}{\sqrt{3}}}cos(k_yA_0).
\end{eqnarray}
Whereas the fourth NN hopping terms are given by,
\begin{eqnarray}
h^{AB}_{11}(\mathbf{k})&=&u^{22}_{11}e^{\frac{i5k_xA_0}{2\sqrt{3}}}+\frac{1}{2}(u^{22}_{11}+3u^{22}_{22})[e^{-\frac{2ik_xA_0}{\sqrt{3}}}cos({k_yA_0})+e^{-\frac{ik_xA_0}{2\sqrt{3}}}cos\frac{3}{2}k_yA_0],\\
h^{AB}_{12/21}(\mathbf{k})&=& -\frac{\sqrt{3}i}{2}(u^{22}_{11}-u^{22}_{22})[e^{-\frac{2ik_xA_0}{\sqrt{3}}}sin(\frac{k_y A_0}{2})+e^{-\frac{ik_xA_0}{2\sqrt{3}}}sin\frac{3}{2}k_y],\\
h^{AB}_{22}(\mathbf{k})&=&u^{22}_{22}e^{\frac{5ik_xA_0}{2\sqrt{3}}}+\frac{1}{2}(3u^{22}_{11}+u^{22}_{22})[e^{-\frac{2ik_xA_0}{\sqrt{3}}}cos({k_yA_0})+e^{-\frac{ik_xA_0}{2\sqrt{3}}}cos\frac{3}{2}k_yA_0].
\end{eqnarray}
Finally, the fifth NN hopping terms are,
\begin{eqnarray}
h^{AA/BB}_{11}(\mathbf{k})&=&2q^{22}_{11}cos(\sqrt{3}k_x A_0)+(q^{22}_{11}+3q^{22}_{22})cos\frac{3}{2}k_y A_0cos\frac{\sqrt{3}}{2}k_x A_0,\\
h^{AA/BB}_{22}(\mathbf{k})&=&2q^{22}_{22}cos(\sqrt{3}k_x A_0)+(3q^{22}_{11}+q^{22}_{22})cos\frac{3}{2}k_y A_0cos\frac{\sqrt{3}}{2}k_x A_0,\\
h^{AA/BB}_{12/21}(\mathbf{k})&=&\sqrt{3}(q^{22}_{11}-q^{22}_{22})sin\frac{3}{2}k_y A_0sin\frac{\sqrt{3}}{2}k_y A_0\pm 2iq^{22}_{12}[sin(\sqrt{3}k_x A_0)-2cos(\frac{3}{2}k_y a_0)sin(\frac{\sqrt{3}}{2}k_x A_0)].
\label{5NNpxpy}
\end{eqnarray}
Here $A_0$ is the inplane lattice constant for the superlattice. In the absence of intervalley coupling, the hopping parameters satisfy $s^{22}_{11}=s^{22}_{22}$ ($s=t,T,p,u,q$). Therefore, there are six independent hopping parameters. According to our fitting to the band structure from the continuum model taken from Ref.\onlinecite{Koshino2018PRX}, the parameters for the band structure as given in the main text are (in meV),
\begin{eqnarray}
&&t^{22}_{11}=0.5269 \quad T^{22}_{11}=0.0043 \quad p^{22}_{11}=0.0743    \nonumber\\
&& u^{22}_{11}=    0.1147 \quad  q^{22}_{11}=   -0.0059 \quad q^{22}_{12}= 0.1006.
\end{eqnarray}
The obtained band structure is shown in Fig.\ref{bandcom} in comparison with that from the continuum model. Despite narrower bandwidth and quantitative differences around the $\Gamma$ point, this model captures the main features of band structure from the continuum model. An intervalley coupling can be introduced by $s^{22}_{11}\neq s^{22}_{22}$ ($s=t,T,p,u,q$).

\begin{figure}[tb]
\centerline{\includegraphics[width=0.8\columnwidth]{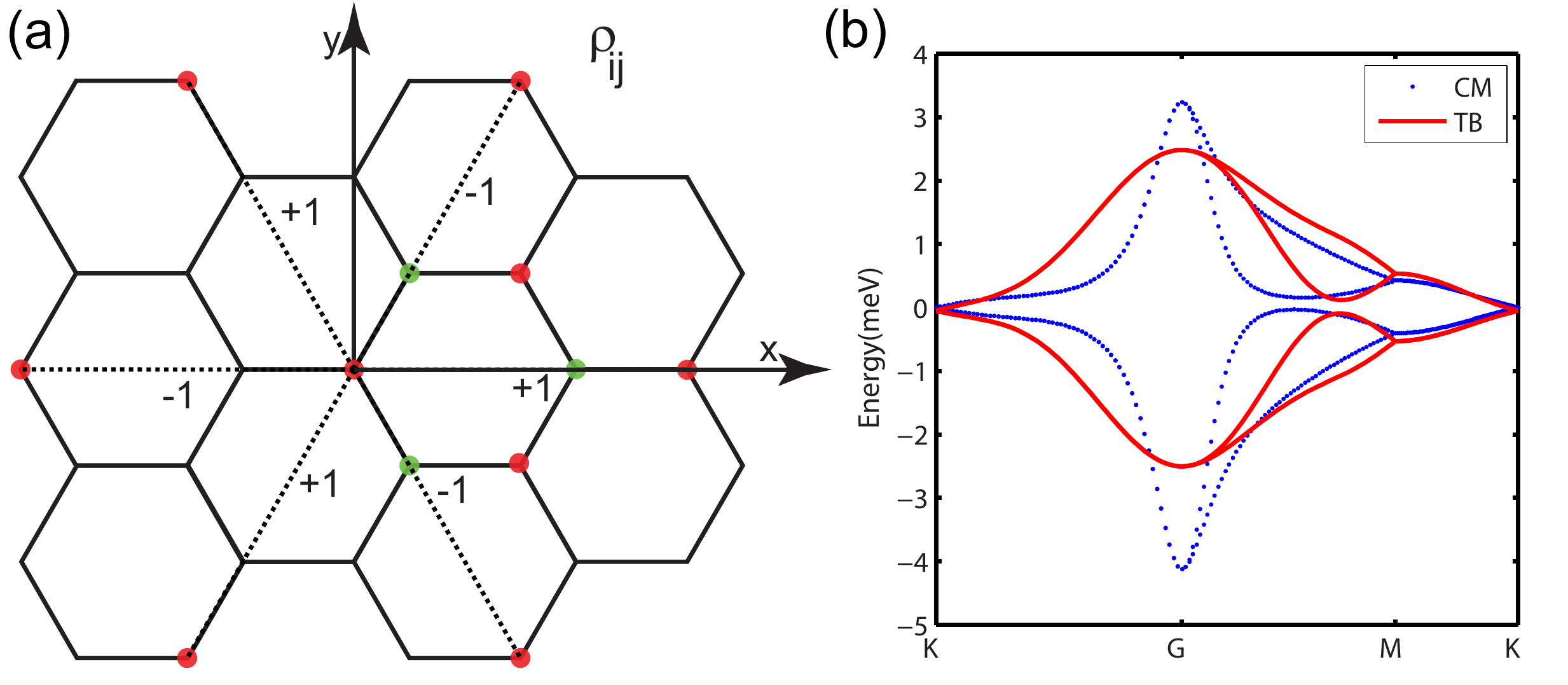}}
\caption{(color online) (a) The factor $\rho_{ij}$ for the 5th NN hopping terms. (b) Band structure from the continuum model (CM, blue dot) with parameters taken from Ref.\onlinecite{Koshino2018PRX} and the effective tight-binding model including up to 5-th NN hoppings (TB, red lines).    \label{bandcom}  }
\end{figure}

\section{RPA method} \label{rpamethod}
The adopted interactions are given in the main text. The bare susceptibility is defined as,
\begin{eqnarray}
\chi^{0}_{l_1l_2l_3l_4}(\bm{q},\tau)=\frac{1}{N}\sum_{\bm{k}\bm{k}'}\langle T_{\tau} c^{\dag}_{l_3\sigma}(\bm{k}+\bm{q},\tau)c_{l_4\sigma}(\bm{k},\tau)c^{\dag}_{l_2\sigma}(\bm{k}'-\bm{q},0)c_{l_1\sigma}(\bm{k}',0) \rangle_0 .
\end{eqnarray}
where $l_i$ is the orbital indices. The bare susceptibility in momentum-frequency is given by,
\begin{eqnarray}
\chi^0_{l_1l_2l_3l_4}(\bm{q},i\omega_n)\!\!=\!\!-\frac{1}{N}\!\!\sum_{k\mu\nu}a^{l_4}_\mu(\bm{k})a^{l_2*}_{\mu}(\bm{k}) a^{l_1}_\nu(\bm{k}+\bm{q})a^{l3*}_{\nu}(\bm{k}+\bm{q})\frac{n_F(E_{\mu}(\bm{k}))-n_F(E_{\nu}(\bm{k}+\bm{q}))}{i\omega_n+E_{\mu}(\bm{k})-E_{\nu}(\bm{k}+\bm{q})}.
\end{eqnarray}
where $\mu/\nu$ is the band index, $n_F(\epsilon)$ is the Fermi distribution function, $a^{l_i}_\mu(\bm{k})$ is the $l_i$-th component of the eigenvector for band $\mu$ resulting from the diagonalization of the initial Hamiltonian $H_0$ and $E_{\mu}(\bf{k})$ is the eigenvalue of band $\mu$.

The eigenvalues of the bare susceptibility matrix $\chi^0_{l_1l_1;l_2l_2}(\bm{q},i\omega_n)$ for TBG are given in Fig.\ref{susbzNTBocc17}, which show the intrinsic spin fluctuations in the system.
The interacting spin susceptibility and the charge susceptibility in RPA are given by,
\begin{eqnarray}
\chi^{RPA}_1(\bm{q})&=&[1-\chi_0(\bm{q})U^s(\bm{q})]^{-1}\chi_0(\bm{q}),\\
\chi^{RPA}_0(\bm{q})&=&[1+\chi_0(\bm{q})U^c(\bm{q})]^{-1}\chi_0(\bm{q}),
\label{RPA1}
\end{eqnarray}
where $U^s$, $U^c$ are the interaction matrices are defined as,
\begin{eqnarray}
U^s_{\alpha l_1,\alpha l_2;\alpha l_3,\alpha,l_4}(\bm{q})&=&
\begin{cases}
U+U_{2,\alpha\alpha}(\bm{q})   & l_1=l_2=l_3=l_4,\\
U'+U'_{2,\alpha\alpha}(\bm{q})    & l_1=l_3\neq l_2=l_4,\\
J   & l_1=l_2\neq l_3=l_4,\\
J'   & l_1=l_4\neq l_2=l_3,\\
\end{cases}\\
\label{EQ:Us}
U^c_{\alpha l_1,\alpha l_2; \alpha l_3,\alpha l_4}(\bm{q})&=&
\begin{cases}
U+U_{2,\alpha\alpha}(\bm{q})   & l_1=l_2=l_3=l_4,\\
-U'+2J-U'_{2,\alpha\alpha}(\bm{q})   & l_1=l_3\neq l_2=l_4,\\
2U'-J+2U'_{2,\alpha\alpha}(\bm{q})   & l_1=l_2\neq l_3=l_4,\\
J'   & l_1=l_4\neq l_2=l_3,\\
\end{cases}\\
\label{EQ:Uc1}
U^c_{\alpha l_1,\alpha l_2; \bar{\alpha} l_3,\bar{\alpha} l_4}(\bm{q})&=&
\begin{cases}
2U_{1,\alpha\bar{\alpha}}(\bm{q})+2U_{3,\alpha\bar{\alpha}}(\bm{q})   & l_1=l_2=l_3=l_4,\\
2U'_{1,\alpha\bar{\alpha}}(\bm{q})+2U'_{3,\alpha\bar{\alpha}}(\bm{q})  & l_1=l_2\neq l_3=l_4.\\
\end{cases}
\label{EQ:Uc2}
\end{eqnarray}
Here $\alpha$ is the sublattice index. The nonlocal interaction are $U_{n,\alpha\beta}(\bm{q})=U_{n,\alpha\beta}f^{\alpha\beta}_n(\bm{q})$ and $U'_{n,\alpha\beta}(\bm{q})=U'_{n,\alpha\beta}f^{\alpha\beta}_n(\bm{q})$, with the form factors
\begin{eqnarray}
f^{AB}_1(\bm{q})&=&e^{\frac{iq_xA_0}{\sqrt{3}}}+2e^{-\frac{iq_xA_0}{2\sqrt{3}}}cos\frac{q_yA_0}{2},\\
f^{\alpha\alpha}_2(\bm{q})&=&2cos(q_y A_0)+4cos\frac{\sqrt{3}}{2}q_x A_0cos\frac{1}{2}q_yA_0, \\
f^{AB}_3(\bm{q})&=&e^{-\frac{2iq_xA_0}{\sqrt{3}}}+2e^{\frac{iq_xA_0}{\sqrt{3}}}cosq_yA_0,
\end{eqnarray}
and $f^{BA}_{1,3}(\bm{q})=f^{AB}_{1,3}(-\bm{q})$.

The effective interaction obtained in the RPA approximation is,
\begin{eqnarray} V_{eff}=\sum_{ij,\textbf{k}\textbf{k}'}\Gamma_{ij}(\textbf{k},\textbf{k}')c^{\dag}_{i\textbf{k}\uparrow}c^{\dag}_{i-\textbf{k}\downarrow}c_{j-\textbf{k}'\downarrow}c_{j\textbf{k}'\uparrow}
\end{eqnarray}
where the momenta $\textbf{k}$ and $\textbf{k}'$ are restricted to  different FS $C_i$ with $\textbf{k}\in C_i$ and $\textbf{k}'\in C_j$ and $\Gamma_{ij}(\textbf{k},\textbf{k}')$ is the pairing scattering vertex in the singlet channel\cite{Kemper2010}. The pairing vertex $\Gamma$ is,
\begin{eqnarray}
\Gamma_{ij}(\textbf{k},\textbf{k}')=\sum_{l_1 l_2 l_3 l4}a^{l_2,*}_{v_i}(\textbf{k}) a^{l_3,*}_{v_i}(-\textbf{k}) Re[\Gamma_{l_1 l_2 l_3 l_4}(\textbf{k},\textbf{k}',\omega=0)] a^{l_1}_{v_j}(\textbf{k}') a^{l_4}_{v_j}(-\textbf{k}'),
\end{eqnarray}
 where $a^{l}_{v}$(orbital index $l$ and band index $v$) is the component of the eigenvectors from the diagonalization of the tight-binding Hamiltonian. The orbital vertex function $\Gamma_{l_1 l_2 l_3 l_4}$ for the singlet channel and triplet channel in the fluctuation exchange formulation\cite{Bickers1989,Kubo2007,Kemper2010,Wu2014,Wu2015} are given by,
 \begin{eqnarray}
\Gamma^S_{l_1 l_2 l_3 l_4}(\textbf{k},\textbf{k}',\omega)&=&[\frac{3}{2}\bar{U}^s \chi^{RPA}_1(\textbf{k}-\textbf{k}',\omega)\bar{U}^s+\frac{1}{2}\bar{U}^s -\frac{1}{2}\bar{U}^c\chi^{RPA}_0(\textbf{k}-\textbf{k}',\omega)\bar{U}^c+\frac{1}{2}\bar{U}^c]_{l_1 l_2 l_3 l_4},\\
\Gamma^T_{l_1 l_2 l_3 l_4}(\textbf{k},\textbf{k}',\omega)&=&[-\frac{1}{2}\bar{U}^s \chi^{RPA}_1(\textbf{k}-\textbf{k}',\omega)\bar{U}^s+\frac{1}{2}\bar{U}^s -\frac{1}{2}\bar{U}^c\chi^{RPA}_0(\textbf{k}-\textbf{k}',\omega)\bar{U}^c+\frac{1}{2}\bar{U}^c]_{l_1 l_2 l_3 l_4},
\end{eqnarray}
where $\bar{U}^{s/c}=U^{s/c}(\bm{k}-\bm{k}')$.
The $\chi^{RPA}_0$ describes here the charge fluctuation contribution and the $\chi^{RPA}_1$ the spin fluctuation contribution. For a given gap function $\Delta(\textbf{k})$, the pairing strength functional is,
\begin{eqnarray}
\lambda[\Delta(\textbf{k})]=-\frac{\sum_{ij}\oint_{C_i} \frac{dk_{\|}}{v_F(\textbf{k})} \oint_{C_j} \frac{dk'_{\|}}{v_F(\textbf{k}')} \Delta(\textbf{k})\Gamma_{ij}(\textbf{k},\textbf{k}') \Delta(\textbf{k}')} {4\pi^2\sum_i\oint_{C_i} \frac{dk_{\|}}{v_F(\textbf{k})} [\Delta(\textbf{k})]^2 },
\label{strength}
\end{eqnarray}
where $v_F(\textbf{k})=|\triangledown_{\textbf{k}}E_i(\textbf{k})|$ is the Fermi velocity on a given fermi surface sheet $C_i$. From the stationary condition we find the following eigenvalue problem,
\begin{eqnarray}
-\sum_{j} \oint_{C_j} \frac{dk'_{\|}}{4\pi^2v_F(\textbf{k}')} \Gamma_{ij}(\textbf{k},\textbf{k}') \Delta_{\alpha}(\textbf{k}')=\lambda_{\alpha}\Delta_{\alpha}(\textbf{k}),
\label{strength1}
\end{eqnarray}
where the interaction $\Gamma_{ij}$ is the symmetric (antisymmetric) part of the full interaction in singlet (triplet) channel. The leading eigenfunction $\Delta_{\alpha}(\bf{k})$ and eigenvalue $\lambda_{\alpha}$ are obtained from the above equation. The obtained gap function should have the symmetry of one of the irreducible representations for the corresponding point group.


\section{gap functions for leading states } \label{gapfunction}
The gap functions for the leading states with $U=1.5$ meV and $J/U=0$ are shown in Fig.\ref{gapuj0}. All the pairing state are mainly due to the Fermi surface nesting $\bm{q}_2$. The order parameter connected by $\bm{q}_2$ should have a sign change (the same sign) in single (triplet) channel. For the typical Hund's rule coupling $J/U=0.225$, the two leading states are $A'_{1u}$ and $E_g$ states and their gap functions are displayed in Fig.\ref{gapuj0225}. The gap size of $A'_{1u}$ state is isotropic, mainly attributed to the onsite orbital singlet pairing. The $d$-wave state are similar to those in the case of $J/U=0$ but the gap size of outer pocket gets enhanced.

\begin{figure}[tb]
\centerline{\includegraphics[width=0.8\columnwidth]{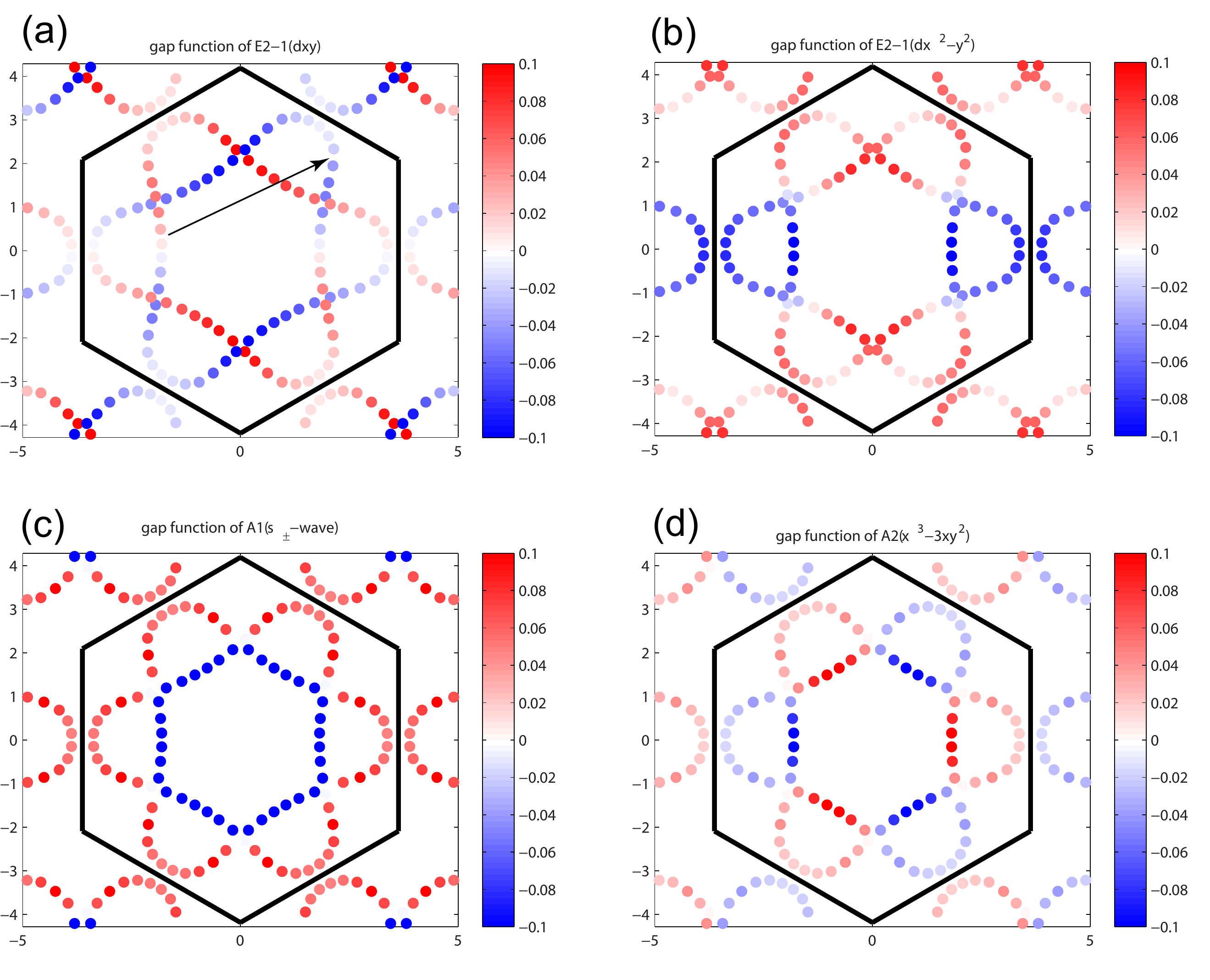}}
\caption{(color online) Gap functions for the three leading pairing states at $n=1.7$ for $U=1.5$ meV and $J/U=0$. (a) $d_{xy}$-wave state (b)  $d_{x^2-y^2}$-wave state, (c) $s_{\pm}$-wave state (d)  $A_{2u}$ ($f_{x^3-3xy^2}$)-wave state.    \label{gapuj0}  }
\end{figure}

\begin{figure}[tb]
\centerline{\includegraphics[width=1\columnwidth]{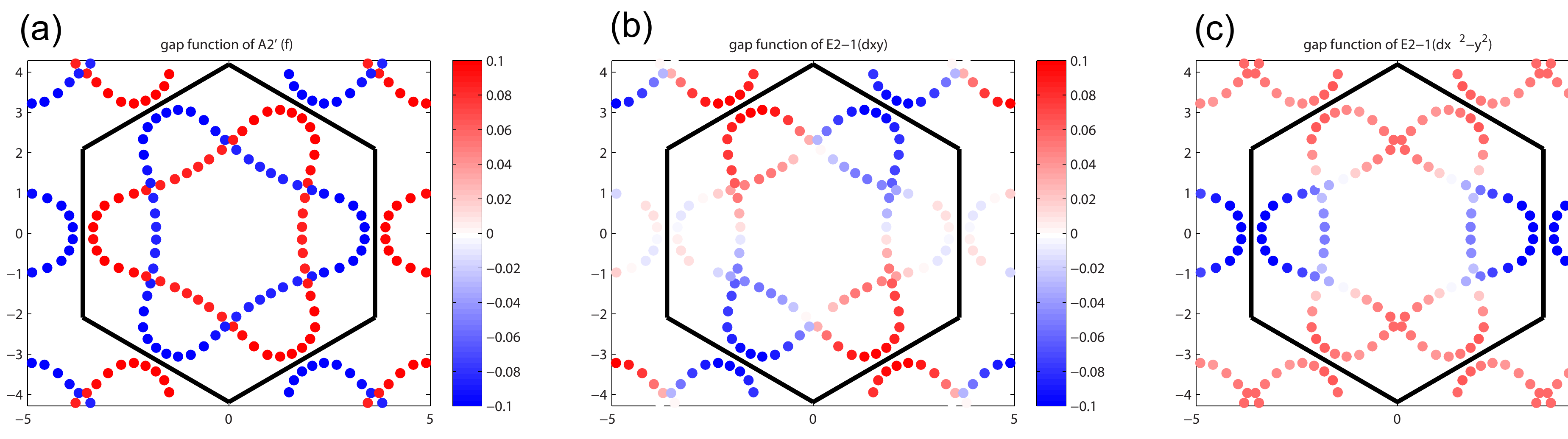}}
\caption{(color online) Gap functions for the two leading pairing states at $n=1.7$ for $U=1.5$ meV and $J/U=0.225$. (a) $A'_{2u}$ state (b) $d_{xy}$-wave state (c)  $d_{x^2-y^2}$-wave state.   \label{gapuj0225}  }
\end{figure}

\section{Real-space pairing with different nonlocal interactions}\label{fitting}

\subsection{classification}
In order to further understand the obtained leading pairing states, we consider the general pairing states in real space in the honeycomb lattice with two orbitals. By defining $\psi^\dag_{\bm{k}\sigma}=(c^{\dag}_{\bm{k}Ax\sigma},c^{\dag}_{\bm{k}Ay\sigma},c^{\dag}_{\bm{k}Bx\sigma},c^{\dag}_{\bm{k}By\sigma})$.
Let $\Psi^\dag_{\bm{k}}=(\psi^\dag_{\bm{k}\uparrow},\psi^\dag_{\bm{k}\downarrow})$, we can write down the pairing state according to the point-group symmetry. Generally, the pairing state can be written as,
\begin{eqnarray}
\hat{\Delta}_\alpha=\sum_{ijln}f^n_{ijl}(\bm{k})\Psi^\dag_{\bm{k}}s_{i}\otimes \sigma_j \otimes \tau_l [\Psi^\dag_{-\bm{k}}]^T,
\end{eqnarray}
where $s$, $\sigma$, $\tau$ are Pauli matrices defined in the spin, sublattice and orbital space. Generally the classification for a multi-orbital system on a lattice can be done in three step: (1) write down the irreducible lattice harmonics according the point group for the $n$-th NN bond\cite{Platt2013}; (2) write down all possible pairings in orbital, spin and sublattice space and classify them using group theory. Each pairing has an IR label; (3) multiply the above two terms to form an IR of lattice point group with further considering the fermionic antisymmetry. These pairings are the HFP. The pairing in orbital space for TBG is provided in the main text and the spin singlet and sin triplet pairing can be represented $s$ matrix,
 \begin{eqnarray}
&S=0& |\uparrow\downarrow-\downarrow\uparrow \rangle,\quad is_2\\
&S=1&
\begin{cases}
 |\uparrow\downarrow+\downarrow\uparrow \rangle,\quad s_1 \\
 |\uparrow\uparrow\rangle,\quad \frac{1}{2}(s_0+s_3) \\
 |\downarrow\downarrow\rangle,\quad \frac{1}{2}(s_0-s_3).
\end{cases}
 \end{eqnarray}
Due to spin rotation invariance for TBG, we only consider the spin-singlet and spin-triplet states both present in the $S_z=0$ pairing sector, i.e. the former two terms. In the following, we provide the lattice harmonics and pairing state up to NNN bond.

For the NN and TNN bond, the pairing happens between two A and B sublattices. The corresponding lattice harmonics are given by,
\begin{eqnarray}
NN: \quad A_1 &&\quad f^1_{A_1}(\bm{k})=\frac{1}{\sqrt{3}}(e^{\frac{ik_x}{\sqrt{3}}}+2e^{-\frac{ik_x}{2\sqrt{3}}}cosk_y/2), \\
E_1 &&\quad f^1_{E_1}(\bm{k})=\sqrt{2}e^{-\frac{ik_x}{2\sqrt{3}}}sink_y/2 \sim \frac{1}{\sqrt{2}}k_y,\\
E_2 &&\quad f^1_{E_2}(\bm{k})=\frac{\sqrt{6}}{3}(e^{-\frac{ik_x}{\sqrt{3}}}-e^{\frac{ik_x}{2\sqrt{3}}}cosk_y/2) \sim \frac{1}{\sqrt{2}}ik_x,\\
TNN: \quad A_1 &&\quad f^3_{A_1}(\bm{k})=\frac{1}{\sqrt{3}}(e^{\frac{-2ik_x}{\sqrt{3}}}+2e^{\frac{ik_x}{\sqrt{3}}}cosk_y), \\
E_1 &&\quad f^3_{E_1}(\bm{k})=\sqrt{2}e^{\frac{ik_x}{\sqrt{3}}}sink_y \sim \frac{1}{\sqrt{2}}k_y,\\
E_2 &&\quad f^3_{E_2}(\bm{k})=\frac{\sqrt{6}}{3}(e^{\frac{-2ik_x}{\sqrt{3}}}-e^{\frac{+ik_x}{\sqrt{3}}}cosk_y/2) \sim \frac{1}{\sqrt{2}}ik_x,
\end{eqnarray}
The pairing in sublattice space is $f(\bm{k})\sigma_+ \pm f(-\bm{k})\sigma_-=\left(\begin{array}{cc}  0 & f(\bm{k}) \\ \pm f(-\bm{k})  & 0\\ \end{array}\right)$, which represents the sublattice singlet or triplet states. Now we combine the lattice harmonics and orbital and sublattice pairings to classify the pairing states in real space, where fermionic anti-symmetry should be always preserved. The NN pairing state with form factor $f^k=f^1_{A_1}(k)$ can then be written as,
\begin{eqnarray}
A_1(T)  &&\quad \sum_{\bm{k}}\Psi^\dag_{\bm{k}}is_{2}\otimes (f^k_R\sigma_1-f^k_I\sigma_2) \otimes \tau_0 [\Psi^\dag_{-\bm{k}}]^T, \\
A_2(S) &&\quad \sum_{\bm{k}}\Psi^\dag_{\bm{k}}s_{1}\otimes (if^k_I\sigma_1+if^k_R\sigma_2) \otimes \tau_0 [\Psi^\dag_{-\bm{k}}]^T, \\
A'_1(T) &&\quad \sum_{\bm{k}}\Psi^\dag_{\bm{k}}s_{1}\otimes (f^k_R\sigma_1-f^k_I\sigma_2)\otimes i\tau_2 [\Psi^\dag_{-\bm{k}}]^T, \\
A'_2(S) &&\quad \sum_{\bm{k}}\Psi^\dag_{\bm{k}}is_{2}\otimes (if^k_I\sigma_1+if^k_R\sigma_2)\otimes i\tau_2 [\Psi^\dag_{-\bm{k}}]^T
\end{eqnarray}
where $S$($T$) means sublattice singlet (triplet) and $f^k_R$ ($f^k_I$) is the real (imaginary) part of $f^k$.
The NN pairing state with the lattice harmonic $f^k=f^1_{E1}(k)$ ($f^1_{E1}(-k)=-[f^1_{E1}(k)]^*$) form factor is given by,
\begin{eqnarray}
E_2(T) &&\quad \sum_{\bm{k}}\Psi^\dag_{\bm{k}}is_{2}\otimes (if^k_I\sigma_1+if^k_R\sigma_2) \otimes \tau_0 [\Psi^\dag_{-\bm{k}}]^T, \\
E_1(S) &&\quad \sum_{\bm{k}}\Psi^\dag_{\bm{k}}s_{1}\otimes (f^k_R\sigma_1-f^k_I\sigma_2) \otimes \tau_0 [\Psi^\dag_{-\bm{k}}]^T, \\
E_2'(T) &&\quad \sum_{\bm{k}}\Psi^\dag_{\bm{k}}s_{1}\otimes (if^k_I\sigma_1+if^k_R\sigma_2)\otimes i\tau_2 [\Psi^\dag_{-\bm{k}}]^T, \\
E_1'(S) &&\quad \sum_{\bm{k}}\Psi^\dag_{\bm{k}}is_{2}\otimes (f^k_R\sigma_1-f^k_I\sigma_2)\otimes i\tau_2 [\Psi^\dag_{-\bm{k}}]^T.
\end{eqnarray}
The NN pairing state with lattice harmonic $f^k=f^1_{E_2}(k)$ ($f^1_{E_2}(-k)=[f^1_{E_2}(k)]^*$), on the other hand, is,
\begin{eqnarray}
E_1(T) &&\quad \sum_{\bm{k}}\Psi^\dag_{\bm{k}}is_{2}\otimes (f^k_R\sigma_1-f^k_I\sigma_2) \otimes \tau_0 [\Psi^\dag_{-\bm{k}}]^T, \\
E_2(S) &&\quad \sum_{\bm{k}}\Psi^\dag_{\bm{k}}s_{1}\otimes (if^k_I\sigma_1+if^k_R\sigma_2) \otimes \tau_0 [\Psi^\dag_{-\bm{k}}]^T, \\
E_1'(T) &&\quad \sum_{\bm{k}}\Psi^\dag_{\bm{k}}s_{1}\otimes (f^k_R\sigma_1-f^k_I\sigma_2)\otimes i\tau_2 [\Psi^\dag_{-\bm{k}}]^T, \\
E_2'(S) &&\quad \sum_{\bm{k}}\Psi^\dag_{\bm{k}}is_{2}\otimes (if^k_I\sigma_1+if^k_R\sigma_2)\otimes i\tau_2 [\Psi^\dag_{-\bm{k}}]^T.
\end{eqnarray}
Similarly, we can classify the pairing for the next NN (NNN) bond. For NNN pairing between the same sublattice, the corresponding lattice harmonics are,
\begin{eqnarray}
A_1 &&\quad f^2_{A_1}(k)=\frac{\sqrt{6}}{3}(cosk_y+2cos\frac{\sqrt{3}}{2}k_xcos\frac{1}{2}k_y), \\
A'_1 &&\quad f^2_{A'_1}(k)= \frac{\sqrt{6}}{3}(isink_y-2icos\frac{\sqrt{3}}{2}k_xsin\frac{1}{2}k_y), \\
E_1 &&\quad f^2_{E_1}(k)=\frac{2}{\sqrt{3}}(cosk_y-cos\frac{\sqrt{3}}{2}k_xcos\frac{1}{2}k_y)\sim \frac{\sqrt{3}}{4}(k^2_x-k^2_y),\\
E_2 &&\quad f^2_{E_2}(k)=2sin\frac{\sqrt{3}}{2}k_xsin\frac{1}{2}k_y \sim \frac{\sqrt{3}}{4}2k_xk_y,\\
E'_1 &&\quad f^2_{E'_1}(k)=\frac{2}{\sqrt{3}}(sink_y+cos\frac{\sqrt{3}}{2}k_xsin\frac{1}{2}k_y)\sim \sqrt{3}k_y,\\
E'_2 &&\quad f^2_{E'_2}(k)=2sin\frac{\sqrt{3}}{2}k_xcos\frac{1}{2}k_y\sim \sqrt{3}k_x.
\end{eqnarray}
Now, we combine the above lattice harmonics with the pairing state in orbital space. As the pairing with an opposite sign ($\sigma_3$) on two sublattice will vanish, we have only the pairing term with the same sign on two sublattice. The NNN pairing state with $A_1$ lattice harmonic can be written as,
\begin{eqnarray}
A_1  &&\quad \sum_{\bm{k}}f^2_{A_1}(\bm{k})\Psi^\dag_{\bm{k}}is_{2}\otimes \sigma_0 \otimes \tau_0 [\Psi^\dag_{-\bm{k}}]^T, \\
A'_1 &&\quad \sum_{\bm{k}}f^2_{A_1} \Psi^\dag_{\bm{k}}s_{1}\otimes \sigma_0 \otimes i\tau_2 [\Psi^\dag_{-\bm{k}}]^T.
\end{eqnarray}
The NNN pairing state with $A'_1$ lattice harmonic is given by,
\begin{eqnarray}
A_1  &&\quad \sum_{\bm{k}}f^2_{A'_1}(\bm{k})\Psi^\dag_{\bm{k}}s_{1}\otimes \sigma_0 \otimes \tau_0 [\Psi^\dag_{-\bm{k}}]^T, \\
A'_1 &&\quad \sum_{\bm{k}}f^2_{A'_1} \Psi^\dag_{\bm{k}}is_{2}\otimes \sigma_0 \otimes i\tau_2 [\Psi^\dag_{-\bm{k}}]^T.
\end{eqnarray}
The NNN pairing state with $E_1$ lattice harmonic can be written as,
\begin{eqnarray}
E_1  &&\quad \sum_{\bm{k}}f^2_{E_1}(\bm{k})\Psi^\dag_{\bm{k}}is_{2}\otimes \sigma_0 \otimes \tau_0 [\Psi^\dag_{-\bm{k}}]^T, \\
E_2 &&\quad \sum_{\bm{k}}f^2_{E_1} \Psi^\dag_{\bm{k}}s_{1}\otimes \sigma_0 \otimes i\tau_2 [\Psi^\dag_{-\bm{k}}]^T.
\end{eqnarray}
The NNN pairing state with $E_2$ lattice harmonic is,
\begin{eqnarray}
E_2  &&\quad \sum_{\bm{k}}f^2_{E_2}(\bm{k})\Psi^\dag_{\bm{k}}is_{2}\otimes \sigma_0 \otimes \tau_0 [\Psi^\dag_{-\bm{k}}]^T, \\
E_1 &&\quad \sum_{\bm{k}}f^2_{E_2} \Psi^\dag_{\bm{k}}s_{1}\otimes \sigma_0 \otimes i\tau_2 [\Psi^\dag_{-\bm{k}}]^T.
\end{eqnarray}
The NNN pairing state with $E'_1$ lattice harmonic can be written as,
\begin{eqnarray}
E_1'  &&\quad \sum_{\bm{k}}f^2_{E'_1}(\bm{k})\Psi^\dag_{\bm{k}}s_{1}\otimes \sigma_0 \otimes \tau_0 [\Psi^\dag_{-\bm{k}}]^T, \\
E_2' &&\quad \sum_{\bm{k}}f^2_{E'_1} \Psi^\dag_{\bm{k}}is_{2}\otimes \sigma_0 \otimes i\tau_2 [\Psi^\dag_{-\bm{k}}]^T,
\end{eqnarray}
and the NNN pairing state with $E_2'$ lattice harmonic is,
\begin{eqnarray}
E_2'  &&\quad \sum_{\bm{k}}f^2_{E'_2}(\bm{k})\Psi^\dag_{\bm{k}}s_{1}\otimes \sigma_0 \otimes \tau_0 [\Psi^\dag_{-\bm{k}}]^T, \\
E_1' &&\quad \sum_{\bm{k}}f^2_{E'_2} \Psi^\dag_{\bm{k}}is_{2}\otimes \sigma_0 \otimes i\tau_2 [\Psi^\dag_{-\bm{k}}]^T.
\end{eqnarray}
The pairing states up to TNN in real space are summarized in Table\ref{pairrealspace}. We notice that the IR in orbital space is generally different from that in band space. In the Table, we label some pairing states as the fitting bases for the following real space decomposition.

\begin{table*}[t]
\caption{\label{pairrealspace} Classification of pairing states in real space. The operations are $D(C_{2y})=is_{2}\otimes \sigma_1 \otimes \tau_3$, $D(C_{3z})=e^{-i\frac{\pi}{3}s_3}\otimes \sigma_1 \otimes e^{\frac{2i\pi}{3}\tau_2}$, $\hat{I}=s_0\otimes\sigma_1\otimes\tau_1$, $C_{2x}=IC_{2y}=s_{2}\otimes \sigma_0 \otimes \tau_2$. $f^k_R$ ($f^k_I$) is the real (imaginary) part of the corresponding form factor. $C_{2y,2x}$ ($C^B_{2y,2x}$) is the symmetry operation in orbital (band) space. "+" or "-" in sublattice and orbital space represents triplet ($\alpha\alpha+\beta\beta$) or singlet ($\alpha\beta-\beta\alpha$) with $\alpha,\beta=A,B/p_x,p_y$.  }
\renewcommand{\multirowsetup}{\centering}
\begin{tabular}{cccccccccccccc}
\hline
\hline
real space+f($\bm{k}$) &irrep. & sublattice& orbital & spin& $C_{3z}$ & $C_{2y}$ & $C_{2x}$   & $I$ &$C^B_{2y}$  &$C^B_{2x}$ & band space irrep & Matrix & label\\
   \hline

  \multirow{4}{*}{onsite}  & $A_1$  & intra(+) & intra(+) & - & + & + & +   &  + & + & + & $A_{1g}(s)$ & $is_{2}\sigma_0\tau_0 $ & ($s$,$G_{s1}$)  \\
                           & $A'_1$ & intra(+) & inter(-) & + & + & + & -   & - &  - & + & $A'_{2u}(f)$ & $s_{1}\sigma_0i\tau_2 $ &  ($f_{x^3-3xy^2}$, $G_{W,f1}$) \\
                           & $E_1$ & intra(+) & intra(-) & - &   & + & - &  - & + & + & $E_{g1}(d_{x^2-y^2})$ & $is_{2}\sigma_0\tau_3$ &      \\
                           & $E_2$ & intra(+) & inter(+) & - &   & - & -   & + &  - & -    & $E_{g2}(d_{xy})$ & $is_{2}\sigma_0\tau_1 $ & \\
                            \hline
 \multirow{4}{*}{NN+$A_1$}  & $A_{1T}$  & inter(+) & intra(+) & - & + & + & +   & + & + & + & $A_{1g}(s_{\pm})$  &
                           $is_{2}(f^k_R\sigma_1-f^k_I\sigma_2)\tau_0$  & ($s_{\pm}$, $G^{NN}_{s2}$) \\
                           & $A_{2S}$  & inter(-) & intra(+) & - & + & - & +    &  - & - & + & $A_{2u}(f)$ & $s_{1}(if^k_I\sigma_1+if^k_R\sigma_2)\tau_0$ & ($f_{x^3-3xy^2}$,$G^{NN}_{f1}$)\\
                           & $A'_{1T}$ & inter(+) & inter(-) & + & + & + & -   & - & - & + & $A'_{2u}(f)$ &
                           $s_{1} (f^k_R\sigma_1-f^k_I\sigma_2) i\tau_2$ & ($f_{x^3-3xy^2}$,$G^{NN}_{W,f2}$) \\
                           & $A'_{2S}$ & inter(-) & inter(-) & - & + & + & +   & + & + & + & $A_1(s_{\pm})$ &
                           $is_{2} (if^k_I\sigma_1+if^k_R\sigma_2)i\tau_2$ & ($s_{\pm}$,$G^{NN}_{s3}$)\\
                           \hline
  \multirow{4}{*}{NN+$E_2$}  & $E_{1T}$  & inter(+) & intra(+) & - &   & + & +  & + & + & + & $E_{g1}(d_{x^2-y^2})$ &$is_{2}(f^k_R\sigma_1-f^k_I\sigma_2)\tau_0$ & ($d_{x^2-y^2}$,$G^{NN}_{x^21}$)\\
                           & $E_{2S}$  & inter(-) & intra(+) & - &   & - & +  &  - & - & + & $E_{g2}(p_x)$ & $s_{1}(if^k_I\sigma_1+if^k_R\sigma_2)\tau_0$ & ($p_x$,$G^{NN}_{p_x1}$)\\
                           & $E'_{1T}$ & inter(+) & inter(-) & + &   & + & -  & - & - & + & $E_{u2}(p_x)$ &
                           $s_{1} (f^k_R\sigma_1-f^k_I\sigma_2) i\tau_2$ &  ($p_x$,$G^{NN}_{p_x2}$) \\
                           & $E'_{2S}$ & inter(-) & inter(-) & - &   & + & +  & + & + & + & $E_{g1}(d_{x^2-y^2})$ &
                           $is_{2} (if^k_I\sigma_1+if^k_R\sigma_2)i\tau_2$ & ($d_{x^2-y^2}$,$G^{NN}_{x^22}$)\\
                           \hline
   \multirow{4}{*}{NN+$E_1$}  & $E_{2T}$  & inter(+) & intra(+) & - &   & - & -   & + &  - & - &$E_{g2}(d_{xy})$ &
                           $is_{2}(if^k_I\sigma_1+if^k_R\sigma_2)\tau_0$ & ($d_{xy}$,$G^{NN}_{xy1}$)  \\
                           & $E_{1S}$  & inter(-) & intra(+) & - &   & + & -   &  - & + & - & $E_{u1}(p_y)$ &
                           $s_{1}(f^k_R\sigma_1-f^k_I\sigma_2)\tau_0$ & ($p_y$,$G^{NN}_{p_y1}$)\\
                           & $E'_{2T}$ & inter(+) & inter(-) & + &   & - & +  & - &  + & - &$E_{u1}(p_y)$ &$s_{1}(if^k_I\sigma_1+if^k_R\sigma_2)i\tau_2$ & ($p_y$,$G^{NN}_{p_y2}$)  \\
                           & $E'_{1S}$ & inter(-) & inter(-) & - &   & - & -  & + &  - & - & $E_{g2}(d_{xy})$ &$is_{2}(f^k_R\sigma_1-f^k_I\sigma_2)i\tau_2$ & ($d_{xy}$,$G^{NN}_{xy2}$) \\
                           \hline
 \multirow{2}{*}{NNN+$A_1$}  & $A_1$  & intra(+) & intra(+) & - & + & + & +   & + &  + & + & $A_1(s_{\pm})$ &
                           $f^2_{A_1}(\bm{k})is_{2}\sigma_0\tau_0$ & ($s_{\pm}$,$G^{NNN}_{s4}$)    \\
                           & $A'_1$ & intra(+) & inter(-) & + & + & + & -  & - & - & + & $A_{2u}(f)$  &
                           $f^2_{A_1}(\bm{k})s_{1}\sigma_0i\tau_2$  & ($f$,$G^{NNN}_{f2}$) \\
                           \hline
  \multirow{2}{*}{NNN+$A'_1$}  & $A_1$  & intra(+) & intra(+) & - & + & + & -  & - & + & - & $A_{1u}(f)$ &
                           $f^2_{A'_1}(\bm{k})s_{1}\sigma_0\tau_0$ & $(f_{y^3-3yx^2})$  \\
                           & $A'_1$ & intra(+) & inter(-) & + & + & + & +  & + & - & - &$A_{2g}(I)$  &
                           $f^2_{A'_1}(\bm{k})is_{2}\sigma_0i\tau_2$ & $(I-wave)$\\
                           \hline
 \multirow{2}{*}{NNN+$E_1$}  & $E_1$  & intra(+) & intra(+) & - & + & + & +  & + &   + & + &$E_{g1}(d_{x^2-y^2})$ &
                           $f^2_{E_1}(\bm{k})is_{2}\sigma_0\tau_0$ & ($d_{x^2-y^2}$,$G^{NNN}_{x^23}$)     \\
                           & $E_2$ & intra(+) & inter(-) & + & + & + & -  & - &  - & + &$E_{u2}(p_x)$  &
                           $f^2_{E_1}(\bm{k})s_{1}\sigma_0i\tau_2$ & ($p_x$,$G^{NN}_{p_x3}$)   \\
                           \hline
 \multirow{2}{*}{NNN+$E_2$}  & $E_2$  & intra(+) & intra(+) & - & + & - & -   & + &  - & - & $E_{g2}(d_{xy})$ &
                           $f^2_{E_2}(\bm{k})is_{2}\sigma_0\tau_0$ & ($d_{xy}$,$G^{NNN}_{xy3}$)     \\
                           & $E_1$ & intra(+) & inter(-) & + & + & - & +   & - & + & - & $E_{u1}(p_y)$  &
                           $f^2_{E_2}(\bm{k})s_{1}\sigma_0i\tau_2$ & ($p_y$,$G^{NN}_{p_y3}$)  \\
                           \hline
  \multirow{2}{*}{NNN+$E_1'$}  & $E_1'$  & intra(+) & intra(+) & - & + & + & -   & - & + & - & $E_{u1}(p_y)$ &
                           $f^2_{E_1'}(\bm{k})s_{1}\sigma_0\tau_0$ & ($p_y$,$G^{NN}_{p_y4}$) \\
                           & $E_2'$ & intra(+) & inter(-) & + & + & + & +  & + & - & - & $E_{g2}(d_{xy})$ &
                           $f^2_{E_1'}(\bm{k})is_{2}\sigma_0i\tau_2$ & ($d_{xy}$,$G^{NNN}_{xy4}$) \\
                           \hline
 \multirow{2}{*}{NNN+$E_2'$}  & $E_2'$  & intra(+) & intra(+) & - & + & - & +  & - & - & + & $E_{u2}(p_x)$ &
                           $f^2_{E_2'}(\bm{k})s_{1}\sigma_0\tau_0$ & ($p_x$,$G^{NN}_{p_x4}$)  \\
                           & $E_1'$ & intra(+) & inter(-) & + & + & - & -   & + & + & + & $E_{g1}(d_{x^2-y^2})$ &
                           $f^2_{E_2'}(\bm{k})is_{2}\sigma_0i\tau_2$ & ($d_{x^2-y^2}$,$G^{NNN}_{x^24}$)\\
                           \hline
 \multirow{4}{*}{TNN+$A_1$}  & $A_{1T}$  & inter(+) & intra(+) & - & + & + & +   & + & + & + & $A_{1g}(s_{\pm})$  &
                           $is_{2}(f^k_R\sigma_1-f^k_I\sigma_2)\tau_0$  & ($s_{\pm}$, $G^{NN}_{s5}$) \\
                           & $A_{2S}$  & inter(-) & intra(+) & - & + & - & +    &  - & - & + & $A_{2u}(f)$ & $s_{1}(if^k_I\sigma_1+if^k_R\sigma_2)\tau_0$ & ($f_{x^3-3xy^2}$,$G^{NN}_{f3}$)\\
                           & $A'_{1T}$ & inter(+) & inter(-) & + & + & + & -   & - & - & + & $A'_{2u}(f)$ &
                           $s_{1} (f^k_R\sigma_1-f^k_I\sigma_2) i\tau_2$ & ($f_{x^3-3xy^2}$,$G^{NN}_{W,f3}$) \\
                           & $A'_{2S}$ & inter(-) & inter(-) & - & + & + & +   & + & + & + & $A_{1g}(s_{\pm})$ &
                           $is_{2} (if^k_I\sigma_1+if^k_R\sigma_2)i\tau_2$ & ($s_{\pm}$,$G^{NN}_{s6}$)\\
                           \hline
  \multirow{4}{*}{TNN+$E_2$}  & $E_{1T}$  & inter(+) & intra(+) & - &   & + & +  & + & + & + & $E_{g1}(d_{x^2-y^2})$ &$is_{2}(f^k_R\sigma_1-f^k_I\sigma_2)\tau_0$ & ($d_{x^2-y^2}$,$G^{NN}_{x^24}$)\\
                           & $E_{2S}$  & inter(-) & intra(+) & - &   & - & +  &  - & - & + & $E_{u2}(p_x)$ & $s_{1}(if^k_I\sigma_1+if^k_R\sigma_2)\tau_0$ & ($p_x$,$G^{NN}_{p_x5}$)\\
                           & $E'_{1T}$ & inter(+) & inter(-) & + &   & + & -  & - & - & + & $E_{u2}(p_x)$ &
                           $s_{1} (f^k_R\sigma_1-f^k_I\sigma_2) i\tau_2$ &  ($p_x$,$G^{NN}_{p_x6}$) \\
                           & $E'_{2S}$ & inter(-) & inter(-) & - &   & + & +  & + & + & + & $E_{g1}(d_{x^2-y^2})$ &
                           $is_{2} (if^k_I\sigma_1+if^k_R\sigma_2)i\tau_2$ & ($d_{x^2-y^2}$,$G^{NN}_{x^25}$)\\
                           \hline
   \multirow{4}{*}{TNN+$E_1$}  & $E_{2T}$  & inter(+) & intra(+) & - &   & - & -   & + &  - & - &$E_{g2}(d_{xy})$ &
                           $is_{2}(if^k_I\sigma_1+if^k_R\sigma_2)\tau_0$ & ($d_{xy}$,$G^{NN}_{xy4}$)  \\
                           & $E_{1T}$  & inter(-) & intra(+) & - &   & + & -   &  - & + & - & $E_{u1}(p_y)$ &
                           $s_{1}(f^k_R\sigma_1-f^k_I\sigma_2)\tau_0$ & ($p_y$,$G^{NN}_{p_y5}$)\\
                           & $E'_{2T}$ & inter(+) & inter(-) & + &   & - & +  & - &  + & - &$E_{u1}(p_y)$ &$s_{1}(if^k_I\sigma_1+if^k_R\sigma_2)i\tau_2$ & ($p_y$,$G^{NN}_{p_y6}$)  \\
                           & $E'_{1S}$ & inter(-) & inter(-) & - &   & - & -  & + &  - & - & $E_{u2}(d_{xy})$ &$is_{2}(f^k_R\sigma_1-f^k_I\sigma_2)i\tau_2$ & ($d_{xy}$,$G^{NN}_{xy5}$) \\
                           \hline

 \hline
 \hline
\end{tabular}

\end{table*}

\subsection{ gap functions for leading states as function of $U_1$}\label{gapV}
\begin{figure}[tb]
\centerline{\includegraphics[width=1\columnwidth]{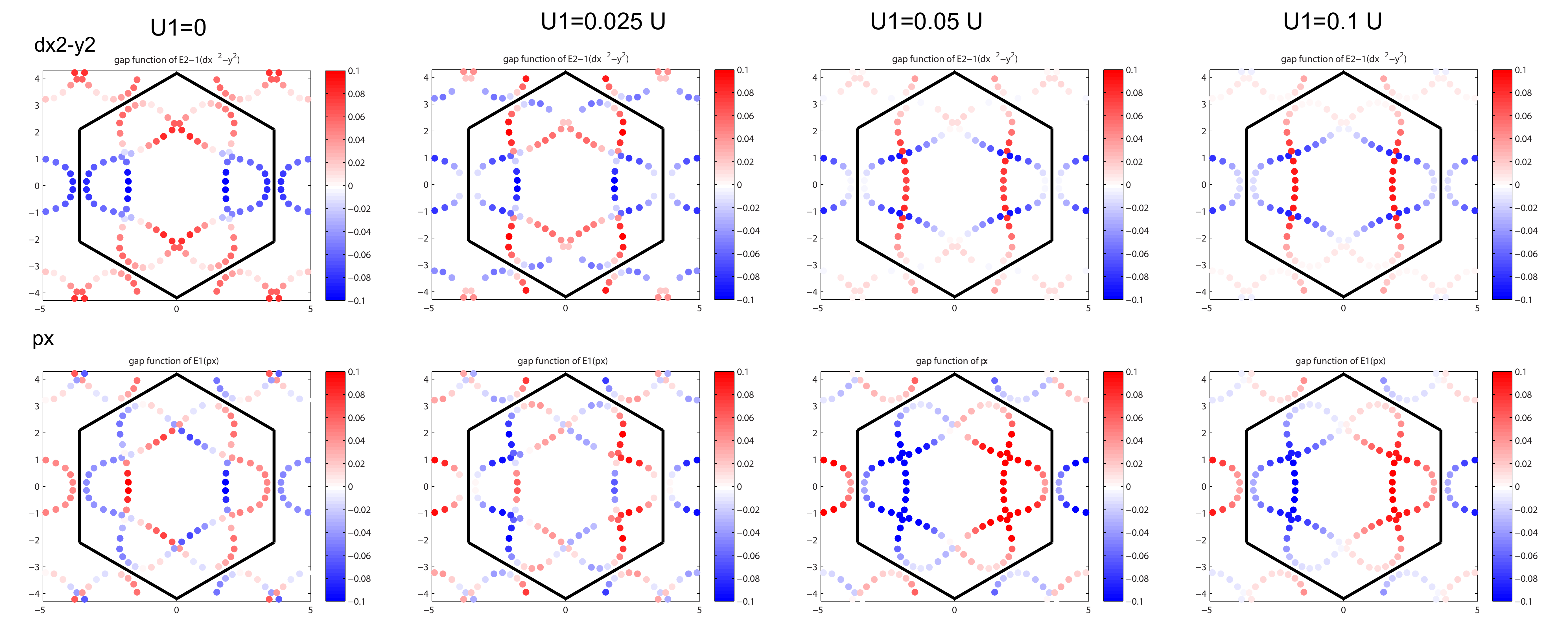}}
\caption{(color online) Gap functions for $d_{x^2-y^2}$ and $p_x$ state as a function of $U_1$ with $n=1.7$ and $J/U=0$. Top panel is for $d_{x^2-y^2}$ state and $p_x$ state.   \label{dxUnn}  }
\end{figure}

\begin{figure}[tb]
\centerline{\includegraphics[width=0.8\columnwidth]{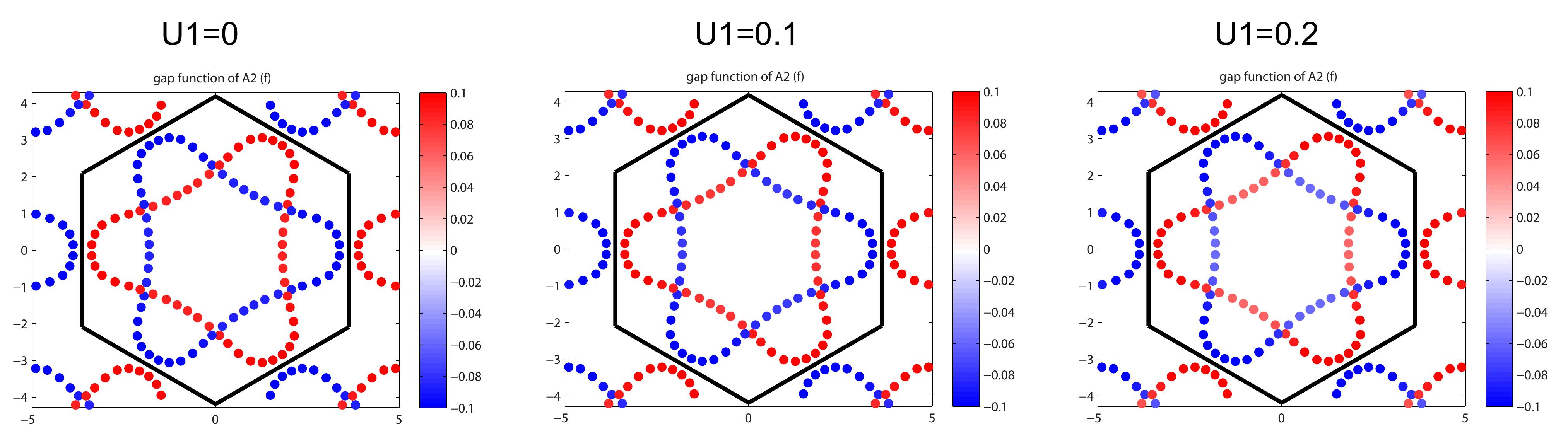}}
\caption{(color online) Gap functions for $A'_{2u}$ state as a function of $U_1$ with $n=1.7$ and $J/U=0.225$. \label{fUnn}  }
\end{figure}
For $J/U=0$, the gap functions of the two leading state ($E_{g}$ and $E_u$) are shown in Fig.\ref{dxUnn}. For both $d_{x^2-y^2}$-wave and $p_x$-wave states, the gap anisotropy increases with increasing nonlocal interaction, which clearly suggests a change in pairing harmonics. For $J/U=0.225$, the gap function for the $A'_{2u}$ state at $U_1=0$ is nearly constant and becomes anisotropic with large $U_1$, as shown in Fig.\ref{fUnn}. It indicates that higher lattice harmonics are more involved.

\subsection{ real-space decomposition for the RPA pairing states with different $U_1$}\label{realpairing}

We, further, project the real-space pairing states onto the Fermi surface and treat them as bases in the decomposition of the obtained gap function from RPA. For a gap function with a certain IR in band space, they can be fitted to the bases with the same IR, where only the first four bases are preserved up to TNN bond. The label for each basis can be found in Table \ref{pairrealspace}.  For $U=0.0015$, $J/U=0$, the $A_{1g}($$s_{\pm}$-wave) pairing state can be decomposed as,
\begin{eqnarray}
\Delta_{s_{\pm}}&=&\alpha_1G_{s1}+\alpha_2 G^{NN}_{s2}+\alpha_3 G^{NN}_{s3}+\alpha_4 G^{NNN}_{s4}\nonumber\\
%
U_1=0.0U \quad &&    0.98 \quad   1.48 \quad   -0.34 \quad   1.66 \nonumber\\
U_1=0.025U \quad &&   1.07 \quad   1.60 \quad  -0.29 \quad   1.67 \nonumber\\
U_1=0.1U \quad &&     1.36 \quad   1.95 \quad   0.03 \quad   1.56 \nonumber\\
U_1=0.2U \quad &&     1.45 \quad   2.16 \quad   -0.07 \quad   1.74 \nonumber\\
\end{eqnarray}
The dominant $E_g$ ($d_{xy}$-wave) pairing state can similarly be decomposed as,
\begin{eqnarray}
\Delta_{d_{xy}}\quad&&\Delta_{d_{xy}}= \beta_1 G^{NN}_{xy1}+\beta_2 G^{NN}_{xy2}+\beta_3 G^{NNN}_{xy3}+\beta_4 G^{NNN}_{xy4}\nonumber\\
  U_1=0.0U \quad &&     0.39 \quad   1.66 \quad   1.62 \quad   1.52  \nonumber\\
 U_1=0.025U \quad &&    0.84 \quad   2.94 \quad   1.66  \quad  2.93  \nonumber\\
 U_1=0.05U \quad &&      -0.01 \quad  0.43 \quad   0.23 \quad   1.36   \nonumber\\
 U_1=0.1U \quad &&      -0.02 \quad  0.11 \quad   0.03 \quad  1.08    \nonumber\\
 U_1=0.2U \quad &&      -0.02 \quad  0.09 \quad   0.03 \quad  1.06    \nonumber\\
\end{eqnarray}

The $p_{x}$-wave pairing state can be decomposed as,
\begin{eqnarray}
\Delta_{p_x}\quad&&\Delta_{p_x}= \gamma_1 G^{NN}_{p_x1}+\gamma_2 G^{NN}_{p_x2}+\gamma_3 G^{NNN}_{p_x3}+\gamma_4 G^{NNN}_{p_x4}\nonumber\\
  U_1=0.0U \quad &&    -2.00 \quad    0.47\quad   -1.66\quad     1.72 \nonumber\\
 U_1=0.025U \quad &&   -2.41 \quad    0.56\quad   -1.56 \quad    2.59 \nonumber\\
 U_1=0.1U \quad &&     0.08 \quad   0.01\quad      -0.03\quad    1.00 \nonumber\\
 U_1=0.2U \quad &&     0.09 \quad   0.02\quad      -0.04 \quad   1.00 \nonumber\\
\end{eqnarray}

The $A_{2u}$($f_{x^3-3xy^2}$-wave) pairing state can be decomposed as,
\begin{eqnarray}
\Delta_{f_{x^3-3xy^2}}\quad&&\Delta_{f}= \eta_1 G^{NN}_{f1}+\eta_2 G^{NNN}_{f2}\nonumber\\
  U_1=0.0U \quad &&      0.55 \quad    1.01  \nonumber\\
 U_1=0.025U \quad &&     0.64 \quad   1.14  \nonumber\\
 U_1=0.1U \quad &&       0.82 \quad   1.40  \nonumber\\
 U_1=0.2U \quad &&       0.70 \quad   1.28  \nonumber\\
\end{eqnarray}
We, further, perform calculations by gradually including nonlocal interactions and then do the real-space decomposition for the dominant $d$-wave pairing states. The obtained coefficients are shown in Table.\ref{orbital-dwave}. We find that NNN interaction $U_2$ suppresses the pairing on NN bond and intraorbital pairing on NNN bond.

For $U=0.0015$, $J/U=0.225$, the $A'_{2u}$( $f$-wave) pairing state can be decomposed as,
\begin{eqnarray}
A'_{2u} \quad &&\Delta_{f}=\eta_1 G_{W,f1}+\eta_2 G^{NN}_{W,f2}+\eta_3 G^{TNN}_{W,f3}\nonumber\\
  U_1=0.0U \quad &&   1.02 \quad   0.15  \quad  0.05 \nonumber\\
 U_1=0.1U \quad &&    1.07  \quad  0.25   \quad 0.11 \nonumber\\
 U_1=0.2U \quad &&    1.015\quad   0.49  \quad  0.21   \nonumber\\
\end{eqnarray}
The $E_{u}$( $p_x$-wave) pairing state can be decomposed as,
\begin{eqnarray}
\Delta_{p_x}\quad&&\Delta_{p_x}= \gamma_1 G^{NN}_{p_x1}+\gamma_2 G^{NN}_{p_x2}+\gamma_3 G^{NNN}_{p_x3}+\gamma_4 G^{NNN}_{p_x4}\nonumber\\
 U_1=0.0U \quad  &&  -1.02 \quad   0.13  \quad  -0.59  \quad  1.04 \nonumber\\
 U_1=0.1U \quad  &&  -0.05 \quad   -0.02 \quad    -0.17  \quad 1.15 \nonumber\\
 U_1=0.2U \quad  &&   0.07  \quad   0.07 \quad   -0.26  \quad  0.99 \nonumber\\
\end{eqnarray}

\begin{table}[t]
\caption{\label{orbital-dwave} Real-space decomposition for $d$-wave state with gradually including long-range interactions. The interaction parameters are $U=1.5$ meV, $U_1=0.05U$ and $U_2=U_3=U_1/2$. NN (NNN) represent pairing on the NN (NNN) bond.   }
\renewcommand{\multirowsetup}{\centering}
\begin{tabular}{ccccc}
\hline
\hline
interaction & NN(intraorbital) & NN(interorbital) & NNN(intraorbital) & NNN(interorbital)\\
   \hline
 $U$ &  0.39 & 1.66 & 1.62 & 1.52 \\
 $U$,$U_1$ & 0.32  & 1.28 & 1.63 &  1.26    \\
 $U$,$U_1$, $U_2$ & 0.05  & 0.63 & 0.33 &  1.53    \\
  $U$,$U_1$, $U_2$, $U_3$ & -0.01 & 0.43 & 0.23 &  1.36    \\
\hline
 \hline
\end{tabular}
\end{table}

\section{microscopic explanation for the pairing states }
\subsection{ onsite interaction }\label{bubbleonsite}
\begin{figure}[tb]
\centerline{\includegraphics[width=0.5\columnwidth]{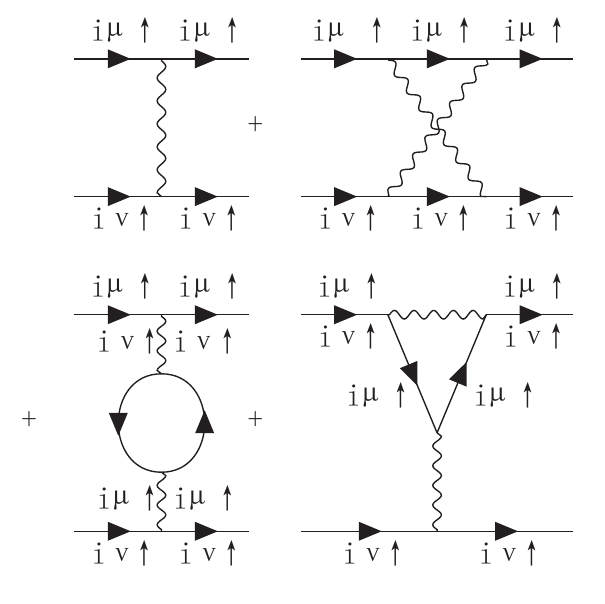}}
\caption{(color online) Effective pairing interaction in the orbital singlet channel up to the second order. (a) $V_1=(U'-J)$, (b) $V_2=(U'-J)^2\chi^{\mu\nu}_{\nu\nu}$, (c) $V_3=-2(U'-J)^2\chi^{\nu\mu}_{\nu\mu}$ (d) $V_4=-2(U'-J)^2\chi^{\nu\mu}_{\nu\mu}$.   \label{onsitediagram}  }
\end{figure}
With only onsite interactions, $d$-wave and $s_{\pm}$-wave pairings are the dominant pairing states, when the Hund's rule coupling is weak ($J/U<0.2$), which is attributed to the Fermi surface nesting at $\bm{q}_1$. However, an onsite spin-triplet pairing will dominate when $J/U$ is greater than 0.2, which is attributed to an orbital-singlet onsite pairing.

 We reveal the mechanism for this pairing in what follows. In Fig.\ref{onsitediagram}, we plot the pairing interaction diagrams up to the second order in the onsite orbital-singlet channel. The first two give repulsive interactions and the latter bubble diagrams attractive interactions. When Hund's rule coupling is small, the first order diagram dominates and the effective interaction is repulsive, therefore orbital-singlet pairing will not be favored. With increasing $J/U$, $\chi(\bm{q}_2)$ get enhanced significantly, as shown in Fig.\ref{susbzNTBocc17}(c). Simultaneously, the first-order repulsive interaction decreases and the contribution from latter bubble diagrams increases rapidly. When $J/U>0.2$, the contribution from the bubble diagram can overcome the first-order term, which gives rise to the orbital-singlet spin-triplet onsite pairing. The superconducting gaps connected by the $\bm{q}_2$ vector should have the same sign due to the effective attractive interaction.

\subsection{including nonlocal interactions}
\begin{figure}[tb]
\centerline{\includegraphics[width=0.5\columnwidth]{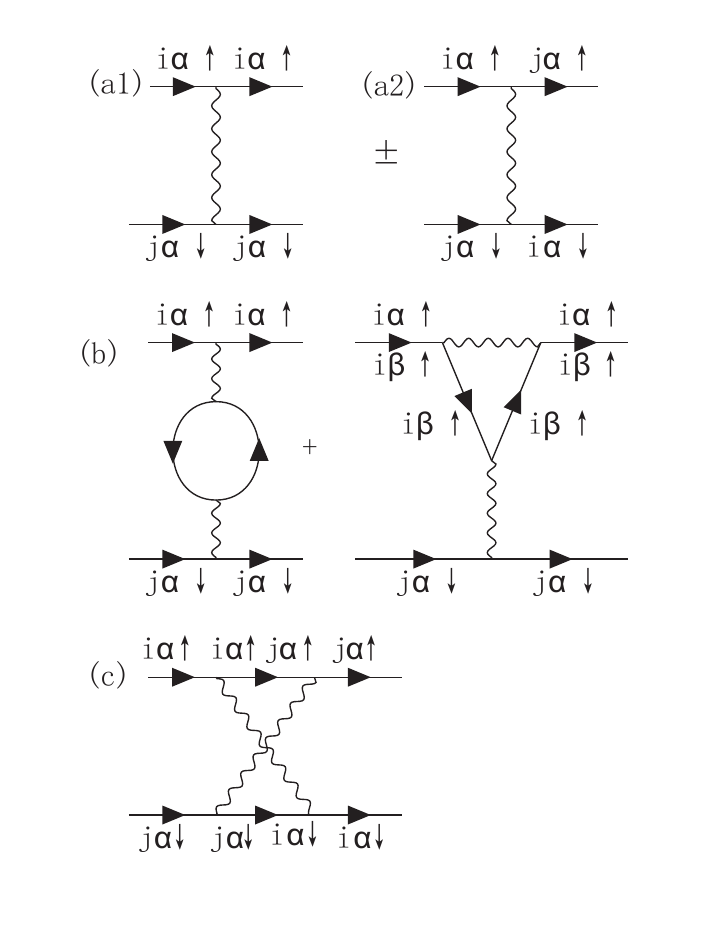}}
\caption{(color online) Effective pairing interaction in spin singlet and triplet channel in honeycomb lattice. $i,j$ is the lattice site and $\alpha$ and $\beta$ are the two different sublattices.   \label{singleeff}  }
\end{figure}
To explain the effect of nonlocal interactions, we start from the single orbital model in the graphene case. For slightly doped graphene, spin-triplet on the same sublattice is found to be dominant when both onsite and nearest-neighbor repulsions are considered\cite{Honerkamp2008}. Due to the Fermonic antisymmetry, the effective vertex function in real space is even(odd) in spin-singlet (triplet) pairing channel, as shown in Fig.\ref{singleeff}(a). The first term in Fig.\ref{singleeff}(a1) can be attractive through the bubble diagrams (see Fig.\ref{singleeff}(b)) while the (a2) term is repulsive, as only the ladder diagram is allowed in the second order (see Fig.\ref{singleeff}(c)). Therefore, the antisymmetric part of the effective interaction is attractive in the spin-triplet channel. This explains the obtained triplet pairing the calculations.

In a multiorbital case, the system has an additional degree of freedom in the orbital space. In our calculations, we find that the system favors the interorbital (orbital-singlet) pairing on the NNN bond for $d$-wave state in the spin-singlet channel and the orbital-triplet (intra-orbital) pairing for the $p$-wave state in the spin-triplet channel, and  will be favored with including nonlocal interactions.

It can be similarly diagrammatically understood as the single orbital case, discussed above. We take the effective vertex function in spin singlet channel as an example. They are even (odd) in space for the orbital-triplet(singlet) channel and given in Fig.\ref{multiorbitaleff}. In our case, the repulsive interactions up to the third NN are included. The first order diagrams contribute repulsive interactions. However, in the second order, the bubble diagram can be attractive (see Fig. \ref{Vsinglet} and \ref{Vtriplet} ). Except the first terms in Fig.\ref{multiorbitaleff} (a) and (b), the other terms can only have ladder diagrams from the onsite and nonlocal interactions in the second order hence are repulsive. Therefore, the vertex function in orbital-singlet channel can be attractive, while in orbital-triplet channel is repulsive. This explains that the $U_2$($U'_2$) will suppress the intraorbital  pairing but enhance the interorbital pairing (orbital-singlet) on the NNN bond for $d$-wave state. Similarly, we can explain the dominant intraorbital pairing for the $p$-wave state in the spin-triplet channel.


\begin{figure}[tb]
\centerline{\includegraphics[width=0.8\columnwidth]{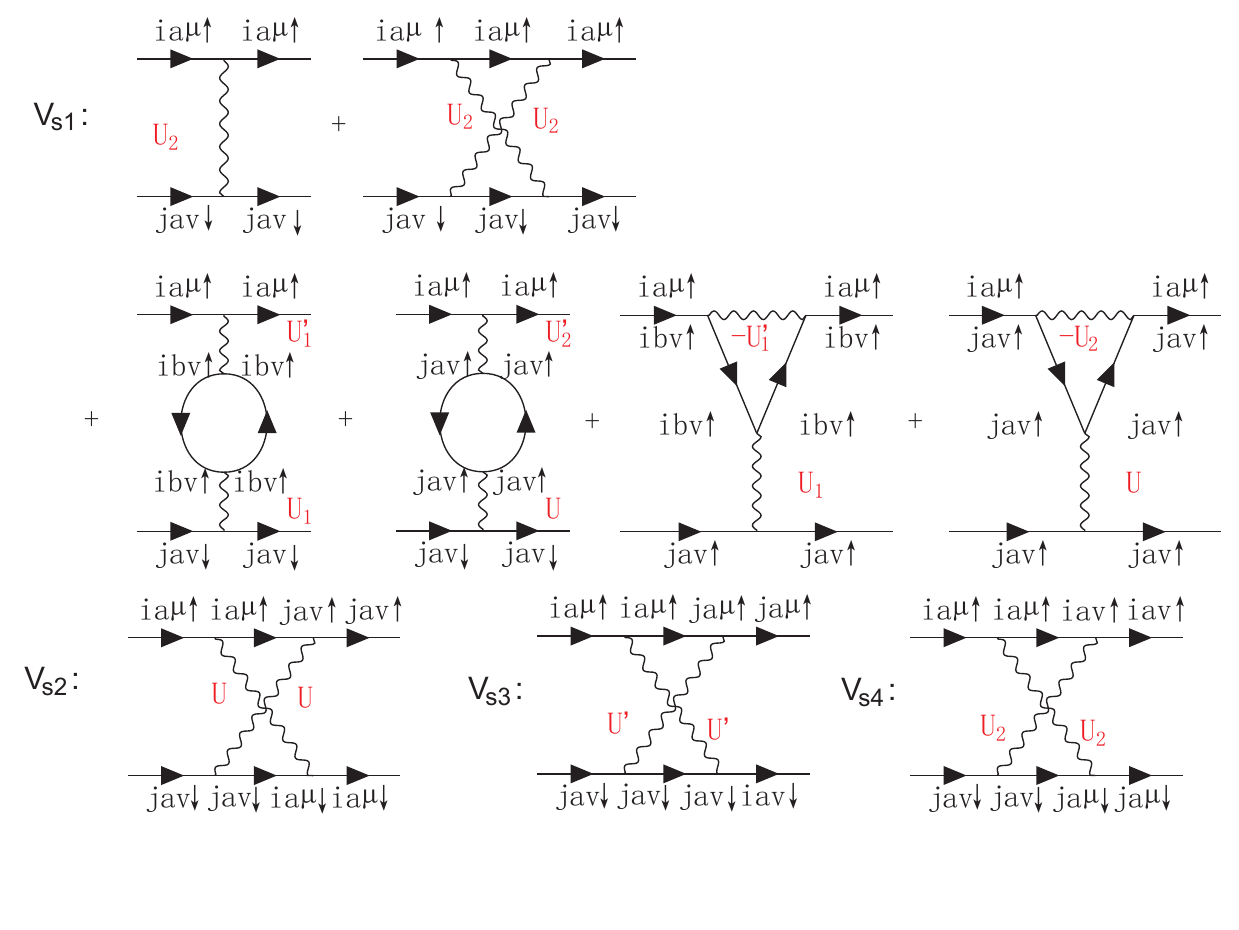}}
\caption{(color online) Effective pairing interaction for the orbital-singlet in the spin-singlet channel up to the second order. $i,j$ denotes the lattice sites and $\mu,\nu$  the orbital indices. \label{Vsinglet}  }
\end{figure}

\begin{figure}[tb]
\centerline{\includegraphics[width=0.8\columnwidth]{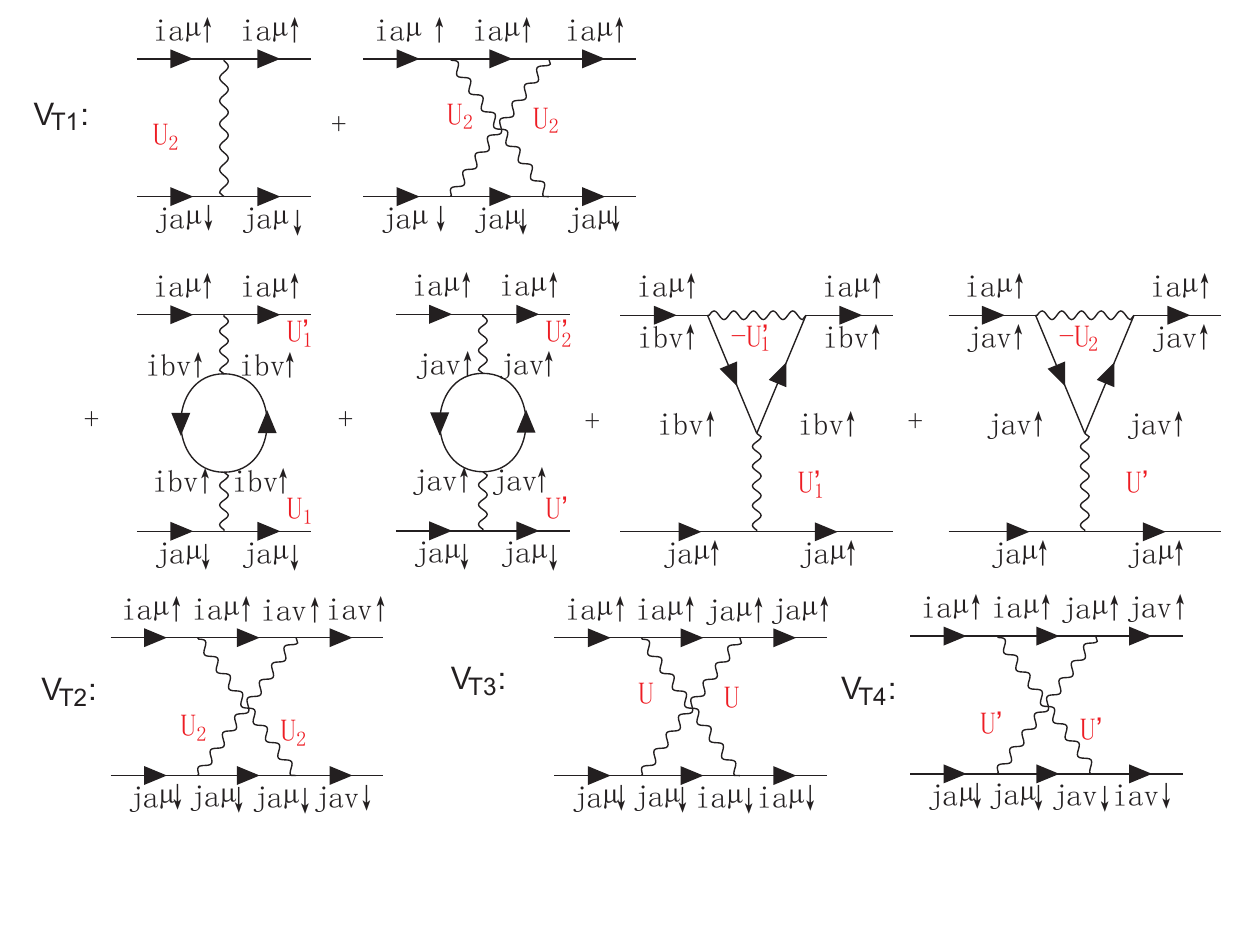}}
\caption{(color online) Effective pairing interaction for the orbital-triplet in the spin singlet channel up to the second order. $i,j$ denotes the lattice sites and $\mu,\nu$ the orbital indices. \label{Vtriplet}  }
\end{figure}

\end{widetext}


\begin{thebibliography}{0}
\expandafter\ifx\csname natexlab\endcsname\relax\def\natexlab#1{#1}\fi
\expandafter\ifx\csname bibnamefont\endcsname\relax
  \def\bibnamefont#1{#1}\fi
\expandafter\ifx\csname bibfnamefont\endcsname\relax
  \def\bibfnamefont#1{#1}\fi
\expandafter\ifx\csname citenamefont\endcsname\relax
  \def\citenamefont#1{#1}\fi
\expandafter\ifx\csname url\endcsname\relax
  \def\url#1{\texttt{#1}}\fi
\expandafter\ifx\csname urlprefix\endcsname\relax\def\urlprefix{URL }\fi
\providecommand{\bibinfo}[2]{#2}
\providecommand{\eprint}[2][]{\url{#2}}

\end{thebibliography}


\begin{thebibliography}{69}
\expandafter\ifx\csname natexlab\endcsname\relax\def\natexlab#1{#1}\fi
\expandafter\ifx\csname bibnamefont\endcsname\relax
  \def\bibnamefont#1{#1}\fi
\expandafter\ifx\csname bibfnamefont\endcsname\relax
  \def\bibfnamefont#1{#1}\fi
\expandafter\ifx\csname citenamefont\endcsname\relax
  \def\citenamefont#1{#1}\fi
\expandafter\ifx\csname url\endcsname\relax
  \def\url#1{\texttt{#1}}\fi
\expandafter\ifx\csname urlprefix\endcsname\relax\def\urlprefix{URL }\fi
\providecommand{\bibinfo}[2]{#2}
\providecommand{\eprint}[2][]{\url{#2}}

\bibitem[{\citenamefont{Cao et~al.}(2018{\natexlab{a}})\citenamefont{Cao,
  Fatemi, Fang, Watanabe, Taniguchi, Kaxiras, and Jarillo-Herrero}}]{Cao2018-1}
\bibinfo{author}{\bibfnamefont{Y.}~\bibnamefont{Cao}},
  \bibinfo{author}{\bibfnamefont{V.}~\bibnamefont{Fatemi}},
  \bibinfo{author}{\bibfnamefont{S.}~\bibnamefont{Fang}},
  \bibinfo{author}{\bibfnamefont{K.}~\bibnamefont{Watanabe}},
  \bibinfo{author}{\bibfnamefont{T.}~\bibnamefont{Taniguchi}},
  \bibinfo{author}{\bibfnamefont{E.}~\bibnamefont{Kaxiras}}, \bibnamefont{and}
  \bibinfo{author}{\bibfnamefont{P.}~\bibnamefont{Jarillo-Herrero}},
  \bibinfo{journal}{Nature} \textbf{\bibinfo{volume}{556}}, \bibinfo{pages}{43}
  (\bibinfo{year}{2018}{\natexlab{a}}).

\bibitem[{\citenamefont{Cao et~al.}(2018{\natexlab{b}})\citenamefont{Cao,
  Fatemi, Demir, Fang, Tomarken, Luo, Sanchez-Yamagishi, Watanabe, Taniguchi,
  Kaxiras et~al.}}]{Cao2018-2}
\bibinfo{author}{\bibfnamefont{Y.}~\bibnamefont{Cao}},
  \bibinfo{author}{\bibfnamefont{V.}~\bibnamefont{Fatemi}},
  \bibinfo{author}{\bibfnamefont{A.}~\bibnamefont{Demir}},
  \bibinfo{author}{\bibfnamefont{S.}~\bibnamefont{Fang}},
  \bibinfo{author}{\bibfnamefont{S.~L.} \bibnamefont{Tomarken}},
  \bibinfo{author}{\bibfnamefont{J.~Y.} \bibnamefont{Luo}},
  \bibinfo{author}{\bibfnamefont{J.~D.} \bibnamefont{Sanchez-Yamagishi}},
  \bibinfo{author}{\bibfnamefont{K.}~\bibnamefont{Watanabe}},
  \bibinfo{author}{\bibfnamefont{T.}~\bibnamefont{Taniguchi}},
  \bibinfo{author}{\bibfnamefont{E.}~\bibnamefont{Kaxiras}},
  \bibnamefont{et~al.}, \bibinfo{journal}{Nature}
  \textbf{\bibinfo{volume}{556}}, \bibinfo{pages}{80}
  (\bibinfo{year}{2018}{\natexlab{b}}).

\bibitem[{\citenamefont{Yankowitz et~al.}(2019)\citenamefont{Yankowitz, Chen,
  Polshyn, Zhang, Watanabe, Taniguchi, Graf, Young, and Dean}}]{Yankowitz2019}
\bibinfo{author}{\bibfnamefont{M.}~\bibnamefont{Yankowitz}},
  \bibinfo{author}{\bibfnamefont{S.}~\bibnamefont{Chen}},
  \bibinfo{author}{\bibfnamefont{H.}~\bibnamefont{Polshyn}},
  \bibinfo{author}{\bibfnamefont{Y.}~\bibnamefont{Zhang}},
  \bibinfo{author}{\bibfnamefont{K.}~\bibnamefont{Watanabe}},
  \bibinfo{author}{\bibfnamefont{T.}~\bibnamefont{Taniguchi}},
  \bibinfo{author}{\bibfnamefont{D.}~\bibnamefont{Graf}},
  \bibinfo{author}{\bibfnamefont{A.~F.} \bibnamefont{Young}}, \bibnamefont{and}
  \bibinfo{author}{\bibfnamefont{C.~R.} \bibnamefont{Dean}},
  \bibinfo{journal}{Science} \textbf{\bibinfo{volume}{363}},
  \bibinfo{pages}{1059} (\bibinfo{year}{2019}).

\bibitem[{\citenamefont{Lu et~al.}(2019)\citenamefont{Lu, Stepanov, Yang, Xie,
  Aamir, Das, Urgell, Watanabe, Taniguchi, Zhang et~al.}}]{LiuXB2019}
\bibinfo{author}{\bibfnamefont{X.}~\bibnamefont{Lu}},
  \bibinfo{author}{\bibfnamefont{P.}~\bibnamefont{Stepanov}},
  \bibinfo{author}{\bibfnamefont{W.}~\bibnamefont{Yang}},
  \bibinfo{author}{\bibfnamefont{M.}~\bibnamefont{Xie}},
  \bibinfo{author}{\bibfnamefont{M.~A.} \bibnamefont{Aamir}},
  \bibinfo{author}{\bibfnamefont{I.}~\bibnamefont{Das}},
  \bibinfo{author}{\bibfnamefont{C.}~\bibnamefont{Urgell}},
  \bibinfo{author}{\bibfnamefont{K.}~\bibnamefont{Watanabe}},
  \bibinfo{author}{\bibfnamefont{T.}~\bibnamefont{Taniguchi}},
  \bibinfo{author}{\bibfnamefont{G.}~\bibnamefont{Zhang}},
  \bibnamefont{et~al.}, \bibinfo{journal}{arXiv e-prints:1903.06513}
  (\bibinfo{year}{2019}).

\bibitem[{\citenamefont{Koshino et~al.}(2018)\citenamefont{Koshino, Yuan,
  Koretsune, Ochi, Kuroki, and Fu}}]{Koshino2018PRX}
\bibinfo{author}{\bibfnamefont{M.}~\bibnamefont{Koshino}},
  \bibinfo{author}{\bibfnamefont{N.~F.~Q.} \bibnamefont{Yuan}},
  \bibinfo{author}{\bibfnamefont{T.}~\bibnamefont{Koretsune}},
  \bibinfo{author}{\bibfnamefont{M.}~\bibnamefont{Ochi}},
  \bibinfo{author}{\bibfnamefont{K.}~\bibnamefont{Kuroki}}, \bibnamefont{and}
  \bibinfo{author}{\bibfnamefont{L.}~\bibnamefont{Fu}},
  \bibinfo{journal}{Physical Review X} \textbf{\bibinfo{volume}{8}},
  \bibinfo{pages}{031087} (\bibinfo{year}{2018}).

\bibitem[{\citenamefont{Po et~al.}(2018)\citenamefont{Po, Zou, Vishwanath, and
  Senthil}}]{Po2018PRX}
\bibinfo{author}{\bibfnamefont{H.~C.} \bibnamefont{Po}},
  \bibinfo{author}{\bibfnamefont{L.}~\bibnamefont{Zou}},
  \bibinfo{author}{\bibfnamefont{A.}~\bibnamefont{Vishwanath}},
  \bibnamefont{and} \bibinfo{author}{\bibfnamefont{T.}~\bibnamefont{Senthil}},
  \bibinfo{journal}{Physical Review X} \textbf{\bibinfo{volume}{8}},
  \bibinfo{pages}{031089} (\bibinfo{year}{2018}).

\bibitem[{\citenamefont{Padhi et~al.}(2018)\citenamefont{Padhi, Setty, and
  Phillips}}]{Padhi2018}
\bibinfo{author}{\bibfnamefont{B.}~\bibnamefont{Padhi}},
  \bibinfo{author}{\bibfnamefont{C.}~\bibnamefont{Setty}}, \bibnamefont{and}
  \bibinfo{author}{\bibfnamefont{P.~W.} \bibnamefont{Phillips}},
  \bibinfo{journal}{Nano Letters} \textbf{\bibinfo{volume}{18}},
  \bibinfo{pages}{6175} (\bibinfo{year}{2018}), ISSN \bibinfo{issn}{1530-6984}.

\bibitem[{\citenamefont{Dodaro et~al.}(2018)\citenamefont{Dodaro, Kivelson,
  Schattner, Sun, and Wang}}]{Dodaro2018}
\bibinfo{author}{\bibfnamefont{J.~F.} \bibnamefont{Dodaro}},
  \bibinfo{author}{\bibfnamefont{S.~A.} \bibnamefont{Kivelson}},
  \bibinfo{author}{\bibfnamefont{Y.}~\bibnamefont{Schattner}},
  \bibinfo{author}{\bibfnamefont{X.~Q.} \bibnamefont{Sun}}, \bibnamefont{and}
  \bibinfo{author}{\bibfnamefont{C.}~\bibnamefont{Wang}},
  \bibinfo{journal}{Physical Review B} \textbf{\bibinfo{volume}{98}},
  \bibinfo{pages}{075154} (\bibinfo{year}{2018}).

\bibitem[{\citenamefont{Liu et~al.}(2018)\citenamefont{Liu, Zhang, Chen, and
  Yang}}]{Liu2108PRL}
\bibinfo{author}{\bibfnamefont{C.-C.} \bibnamefont{Liu}},
  \bibinfo{author}{\bibfnamefont{L.-D.} \bibnamefont{Zhang}},
  \bibinfo{author}{\bibfnamefont{W.-Q.} \bibnamefont{Chen}}, \bibnamefont{and}
  \bibinfo{author}{\bibfnamefont{F.}~\bibnamefont{Yang}},
  \bibinfo{journal}{Physical Review Letters} \textbf{\bibinfo{volume}{121}},
  \bibinfo{pages}{217001} (\bibinfo{year}{2018}).

\bibitem[{\citenamefont{Huang et~al.}(2019)\citenamefont{Huang, Zhang, and
  Ma}}]{Huang2018}
\bibinfo{author}{\bibfnamefont{T.}~\bibnamefont{Huang}},
  \bibinfo{author}{\bibfnamefont{L.}~\bibnamefont{Zhang}}, \bibnamefont{and}
  \bibinfo{author}{\bibfnamefont{T.}~\bibnamefont{Ma}},
  \bibinfo{journal}{Science Bulletin} \textbf{\bibinfo{volume}{64}},
  \bibinfo{pages}{310} (\bibinfo{year}{2019}), ISSN \bibinfo{issn}{2095-9273}.

\bibitem[{\citenamefont{Wu et~al.}(2018{\natexlab{a}})\citenamefont{Wu, Pawlak,
  Jian, and Xu}}]{WuXC2018}
\bibinfo{author}{\bibfnamefont{X.-C.} \bibnamefont{Wu}},
  \bibinfo{author}{\bibfnamefont{K.~A.} \bibnamefont{Pawlak}},
  \bibinfo{author}{\bibfnamefont{C.-M.} \bibnamefont{Jian}}, \bibnamefont{and}
  \bibinfo{author}{\bibfnamefont{C.}~\bibnamefont{Xu}},
  \bibinfo{journal}{arXiv:1805.06906}  (\bibinfo{year}{2018}{\natexlab{a}}).

\bibitem[{\citenamefont{Pizarro et~al.}(2018)\citenamefont{Pizarro, Calder¨®n,
  and Bascones}}]{Pizarro2018}
\bibinfo{author}{\bibfnamefont{J.}~\bibnamefont{Pizarro}},
  \bibinfo{author}{\bibfnamefont{M.}~\bibnamefont{Calder¨®n}},
  \bibnamefont{and} \bibinfo{author}{\bibfnamefont{E.}~\bibnamefont{Bascones}},
  \bibinfo{journal}{arXiv:1805.07303}  (\bibinfo{year}{2018}).

\bibitem[{\citenamefont{Ochi et~al.}(2018)\citenamefont{Ochi, Koshino, and
  Kuroki}}]{Ochi2018}
\bibinfo{author}{\bibfnamefont{M.}~\bibnamefont{Ochi}},
  \bibinfo{author}{\bibfnamefont{M.}~\bibnamefont{Koshino}}, \bibnamefont{and}
  \bibinfo{author}{\bibfnamefont{K.}~\bibnamefont{Kuroki}},
  \bibinfo{journal}{Physical Review B} \textbf{\bibinfo{volume}{98}},
  \bibinfo{pages}{081102} (\bibinfo{year}{2018}).

\bibitem[{\citenamefont{Isobe et~al.}(2018)\citenamefont{Isobe, Yuan, and
  Fu}}]{Isobe2018PRX}
\bibinfo{author}{\bibfnamefont{H.}~\bibnamefont{Isobe}},
  \bibinfo{author}{\bibfnamefont{N.~F.~Q.} \bibnamefont{Yuan}},
  \bibnamefont{and} \bibinfo{author}{\bibfnamefont{L.}~\bibnamefont{Fu}},
  \bibinfo{journal}{Physical Review X} \textbf{\bibinfo{volume}{8}},
  \bibinfo{pages}{041041} (\bibinfo{year}{2018}).

\bibitem[{\citenamefont{Gonz¨¢lez and Stauber}(2019)}]{Gonzalez2018PRL}
\bibinfo{author}{\bibfnamefont{J.}~\bibnamefont{Gonz¨¢lez}} \bibnamefont{and}
  \bibinfo{author}{\bibfnamefont{T.}~\bibnamefont{Stauber}},
  \bibinfo{journal}{Physical Review Letters} \textbf{\bibinfo{volume}{122}},
  \bibinfo{pages}{026801} (\bibinfo{year}{2019}).

\bibitem[{\citenamefont{Kennes et~al.}(2018)\citenamefont{Kennes, Lischner, and
  Karrasch}}]{Kennes2018}
\bibinfo{author}{\bibfnamefont{D.~M.} \bibnamefont{Kennes}},
  \bibinfo{author}{\bibfnamefont{J.}~\bibnamefont{Lischner}}, \bibnamefont{and}
  \bibinfo{author}{\bibfnamefont{C.}~\bibnamefont{Karrasch}},
  \bibinfo{journal}{Physical Review B} \textbf{\bibinfo{volume}{98}},
  \bibinfo{pages}{241407} (\bibinfo{year}{2018}).

\bibitem[{\citenamefont{Kang and Vafek}(2019)}]{Kang2019PRL}
\bibinfo{author}{\bibfnamefont{J.}~\bibnamefont{Kang}} \bibnamefont{and}
  \bibinfo{author}{\bibfnamefont{O.}~\bibnamefont{Vafek}},
  \bibinfo{journal}{Phys. Rev. Lett.} \textbf{\bibinfo{volume}{122}},
  \bibinfo{pages}{246401} (\bibinfo{year}{2019}).

\bibitem[{\citenamefont{Xu and Balents}(2018)}]{XuCK2018PRL}
\bibinfo{author}{\bibfnamefont{C.}~\bibnamefont{Xu}} \bibnamefont{and}
  \bibinfo{author}{\bibfnamefont{L.}~\bibnamefont{Balents}},
  \bibinfo{journal}{Physical Review Letters} \textbf{\bibinfo{volume}{121}},
  \bibinfo{pages}{087001} (\bibinfo{year}{2018}).

\bibitem[{\citenamefont{Fidrysiak et~al.}(2018)\citenamefont{Fidrysiak,
  Zegrodnik, and Spa?ek}}]{Fidrysiak2018}
\bibinfo{author}{\bibfnamefont{M.}~\bibnamefont{Fidrysiak}},
  \bibinfo{author}{\bibfnamefont{M.}~\bibnamefont{Zegrodnik}},
  \bibnamefont{and} \bibinfo{author}{\bibfnamefont{J.}~\bibnamefont{Spa?ek}},
  \bibinfo{journal}{Physical Review B} \textbf{\bibinfo{volume}{98}},
  \bibinfo{pages}{085436} (\bibinfo{year}{2018}).

\bibitem[{\citenamefont{Roy and Juri?i?}(2019)}]{Roy2018}
\bibinfo{author}{\bibfnamefont{B.}~\bibnamefont{Roy}} \bibnamefont{and}
  \bibinfo{author}{\bibfnamefont{V.}~\bibnamefont{Juri?i?}},
  \bibinfo{journal}{Physical Review B} \textbf{\bibinfo{volume}{99}},
  \bibinfo{pages}{121407} (\bibinfo{year}{2019}).

\bibitem[{\citenamefont{You and Vishwanath}(2019)}]{YouYZ2018}
\bibinfo{author}{\bibfnamefont{Y.-Z.} \bibnamefont{You}} \bibnamefont{and}
  \bibinfo{author}{\bibfnamefont{A.}~\bibnamefont{Vishwanath}},
  \bibinfo{journal}{npj Quantum Materials} \textbf{\bibinfo{volume}{4}},
  \bibinfo{pages}{16} (\bibinfo{year}{2019}), ISSN \bibinfo{issn}{2397-4648}.

\bibitem[{\citenamefont{Guo et~al.}(2018)\citenamefont{Guo, Zhu, Feng, and
  Scalettar}}]{Guo2018PRB}
\bibinfo{author}{\bibfnamefont{H.}~\bibnamefont{Guo}},
  \bibinfo{author}{\bibfnamefont{X.}~\bibnamefont{Zhu}},
  \bibinfo{author}{\bibfnamefont{S.}~\bibnamefont{Feng}}, \bibnamefont{and}
  \bibinfo{author}{\bibfnamefont{R.~T.} \bibnamefont{Scalettar}},
  \bibinfo{journal}{Physical Review B} \textbf{\bibinfo{volume}{97}},
  \bibinfo{pages}{235453} (\bibinfo{year}{2018}).

\bibitem[{\citenamefont{Laksono et~al.}(2018)\citenamefont{Laksono, Leaw,
  Reaves, Singh, Wang, Adam, and Gu}}]{Laksono2018}
\bibinfo{author}{\bibfnamefont{E.}~\bibnamefont{Laksono}},
  \bibinfo{author}{\bibfnamefont{J.~N.} \bibnamefont{Leaw}},
  \bibinfo{author}{\bibfnamefont{A.}~\bibnamefont{Reaves}},
  \bibinfo{author}{\bibfnamefont{M.}~\bibnamefont{Singh}},
  \bibinfo{author}{\bibfnamefont{X.}~\bibnamefont{Wang}},
  \bibinfo{author}{\bibfnamefont{S.}~\bibnamefont{Adam}}, \bibnamefont{and}
  \bibinfo{author}{\bibfnamefont{X.}~\bibnamefont{Gu}}, \bibinfo{journal}{Solid
  State Communications} \textbf{\bibinfo{volume}{282}}, \bibinfo{pages}{38}
  (\bibinfo{year}{2018}), ISSN \bibinfo{issn}{0038-1098}.

\bibitem[{\citenamefont{Zhu et~al.}(2018)\citenamefont{Zhu, Xiang, and
  Zhang}}]{ZhuGY2018}
\bibinfo{author}{\bibfnamefont{G.-Y.} \bibnamefont{Zhu}},
  \bibinfo{author}{\bibfnamefont{T.}~\bibnamefont{Xiang}}, \bibnamefont{and}
  \bibinfo{author}{\bibfnamefont{G.-M.} \bibnamefont{Zhang}},
  \bibinfo{journal}{arXiv:1806.07535}  (\bibinfo{year}{2018}).

\bibitem[{\citenamefont{Lin and Tom¨¢nek}(2018)}]{Lin2018PRB}
\bibinfo{author}{\bibfnamefont{X.}~\bibnamefont{Lin}} \bibnamefont{and}
  \bibinfo{author}{\bibfnamefont{D.}~\bibnamefont{Tom¨¢nek}},
  \bibinfo{journal}{Physical Review B} \textbf{\bibinfo{volume}{98}},
  \bibinfo{pages}{081410} (\bibinfo{year}{2018}).

\bibitem[{\citenamefont{Tang et~al.}(2019)\citenamefont{Tang, Yang, Wang,
  Zhang, and Wang}}]{Tang2018}
\bibinfo{author}{\bibfnamefont{Q.-K.} \bibnamefont{Tang}},
  \bibinfo{author}{\bibfnamefont{L.}~\bibnamefont{Yang}},
  \bibinfo{author}{\bibfnamefont{D.}~\bibnamefont{Wang}},
  \bibinfo{author}{\bibfnamefont{F.-C.} \bibnamefont{Zhang}}, \bibnamefont{and}
  \bibinfo{author}{\bibfnamefont{Q.-H.} \bibnamefont{Wang}},
  \bibinfo{journal}{Physical Review B} \textbf{\bibinfo{volume}{99}},
  \bibinfo{pages}{094521} (\bibinfo{year}{2019}).

\bibitem[{\citenamefont{Kozii et~al.}(2019)\citenamefont{Kozii, Isobe,
  Venderbos, and Fu}}]{Kozii2018}
\bibinfo{author}{\bibfnamefont{V.}~\bibnamefont{Kozii}},
  \bibinfo{author}{\bibfnamefont{H.}~\bibnamefont{Isobe}},
  \bibinfo{author}{\bibfnamefont{J.~W.~F.} \bibnamefont{Venderbos}},
  \bibnamefont{and} \bibinfo{author}{\bibfnamefont{L.}~\bibnamefont{Fu}},
  \bibinfo{journal}{Physical Review B} \textbf{\bibinfo{volume}{99}},
  \bibinfo{pages}{144507} (\bibinfo{year}{2019}).

\bibitem[{\citenamefont{Liu et~al.}(2019)\citenamefont{Liu, Li, and
  Yang}}]{LiuZ2018}
\bibinfo{author}{\bibfnamefont{Z.}~\bibnamefont{Liu}},
  \bibinfo{author}{\bibfnamefont{Y.}~\bibnamefont{Li}}, \bibnamefont{and}
  \bibinfo{author}{\bibfnamefont{Y.-F.} \bibnamefont{Yang}},
  \bibinfo{journal}{Chinese Physics B} \textbf{\bibinfo{volume}{28}},
  \bibinfo{pages}{077103} (\bibinfo{year}{2019}), ISSN \bibinfo{issn}{1674-1056
  2058-3834}.

\bibitem[{\citenamefont{Wu et~al.}(2018{\natexlab{b}})\citenamefont{Wu,
  MacDonald, and Martin}}]{WuFC2018PRL}
\bibinfo{author}{\bibfnamefont{F.}~\bibnamefont{Wu}},
  \bibinfo{author}{\bibfnamefont{A.~H.} \bibnamefont{MacDonald}},
  \bibnamefont{and} \bibinfo{author}{\bibfnamefont{I.}~\bibnamefont{Martin}},
  \bibinfo{journal}{Physical Review Letters} \textbf{\bibinfo{volume}{121}},
  \bibinfo{pages}{257001} (\bibinfo{year}{2018}{\natexlab{b}}).

\bibitem[{\citenamefont{Lian et~al.}(2019)\citenamefont{Lian, Wang, and
  Bernevig}}]{Lian2018}
\bibinfo{author}{\bibfnamefont{B.}~\bibnamefont{Lian}},
  \bibinfo{author}{\bibfnamefont{Z.}~\bibnamefont{Wang}}, \bibnamefont{and}
  \bibinfo{author}{\bibfnamefont{B.~A.} \bibnamefont{Bernevig}},
  \bibinfo{journal}{Physical Review Letters} \textbf{\bibinfo{volume}{122}},
  \bibinfo{pages}{257002} (\bibinfo{year}{2019}).

\bibitem[{\citenamefont{Peltonen et~al.}(2018)\citenamefont{Peltonen, Ojaj?rvi,
  and Heikkil?}}]{Peltonen2018}
\bibinfo{author}{\bibfnamefont{T.~J.} \bibnamefont{Peltonen}},
  \bibinfo{author}{\bibfnamefont{R.}~\bibnamefont{Ojaj?rvi}}, \bibnamefont{and}
  \bibinfo{author}{\bibfnamefont{T.~T.} \bibnamefont{Heikkil?}},
  \bibinfo{journal}{Physical Review B} \textbf{\bibinfo{volume}{98}},
  \bibinfo{pages}{220504} (\bibinfo{year}{2018}).

\bibitem[{\citenamefont{Wu et~al.}(2019)\citenamefont{Wu, Hwang, and
  Das~Sarma}}]{WuFC2018-2}
\bibinfo{author}{\bibfnamefont{F.}~\bibnamefont{Wu}},
  \bibinfo{author}{\bibfnamefont{E.}~\bibnamefont{Hwang}}, \bibnamefont{and}
  \bibinfo{author}{\bibfnamefont{S.}~\bibnamefont{Das~Sarma}},
  \bibinfo{journal}{Physical Review B} \textbf{\bibinfo{volume}{99}},
  \bibinfo{pages}{165112} (\bibinfo{year}{2019}).

\bibitem[{\citenamefont{Zou et~al.}(2018)\citenamefont{Zou, Po, Vishwanath, and
  Senthil}}]{Zou2018PRB}
\bibinfo{author}{\bibfnamefont{L.}~\bibnamefont{Zou}},
  \bibinfo{author}{\bibfnamefont{H.~C.} \bibnamefont{Po}},
  \bibinfo{author}{\bibfnamefont{A.}~\bibnamefont{Vishwanath}},
  \bibnamefont{and} \bibinfo{author}{\bibfnamefont{T.}~\bibnamefont{Senthil}},
  \bibinfo{journal}{Physical Review B} \textbf{\bibinfo{volume}{98}},
  \bibinfo{pages}{085435} (\bibinfo{year}{2018}).

\bibitem[{\citenamefont{Song et~al.}(2019)\citenamefont{Song, Wang, Shi, Li,
  Fang, and Bernevig}}]{Song2018}
\bibinfo{author}{\bibfnamefont{Z.}~\bibnamefont{Song}},
  \bibinfo{author}{\bibfnamefont{Z.}~\bibnamefont{Wang}},
  \bibinfo{author}{\bibfnamefont{W.}~\bibnamefont{Shi}},
  \bibinfo{author}{\bibfnamefont{G.}~\bibnamefont{Li}},
  \bibinfo{author}{\bibfnamefont{C.}~\bibnamefont{Fang}}, \bibnamefont{and}
  \bibinfo{author}{\bibfnamefont{B.~A.} \bibnamefont{Bernevig}},
  \bibinfo{journal}{Physical Review Letters} \textbf{\bibinfo{volume}{123}},
  \bibinfo{pages}{036401} (\bibinfo{year}{2019}).

\bibitem[{\citenamefont{Po et~al.}(2019)\citenamefont{Po, Zou, Senthil, and
  Vishwanath}}]{Po2018-2}
\bibinfo{author}{\bibfnamefont{H.~C.} \bibnamefont{Po}},
  \bibinfo{author}{\bibfnamefont{L.}~\bibnamefont{Zou}},
  \bibinfo{author}{\bibfnamefont{T.}~\bibnamefont{Senthil}}, \bibnamefont{and}
  \bibinfo{author}{\bibfnamefont{A.}~\bibnamefont{Vishwanath}},
  \bibinfo{journal}{Physical Review B} \textbf{\bibinfo{volume}{99}},
  \bibinfo{pages}{195455} (\bibinfo{year}{2019}).

\bibitem[{\citenamefont{Carr et~al.}(2019{\natexlab{a}})\citenamefont{Carr,
  Fang, Zhu, and Kaxiras}}]{Carr2019PRR}
\bibinfo{author}{\bibfnamefont{S.}~\bibnamefont{Carr}},
  \bibinfo{author}{\bibfnamefont{S.}~\bibnamefont{Fang}},
  \bibinfo{author}{\bibfnamefont{Z.}~\bibnamefont{Zhu}}, \bibnamefont{and}
  \bibinfo{author}{\bibfnamefont{E.}~\bibnamefont{Kaxiras}},
  \bibinfo{journal}{Physical Review Research} \textbf{\bibinfo{volume}{1}},
  \bibinfo{pages}{013001} (\bibinfo{year}{2019}{\natexlab{a}}).

\bibitem[{\citenamefont{Carr et~al.}(2019{\natexlab{b}})\citenamefont{Carr,
  Fang, Po, Vishwanath, and Kaxiras}}]{Carr2019-2}
\bibinfo{author}{\bibfnamefont{S.}~\bibnamefont{Carr}},
  \bibinfo{author}{\bibfnamefont{S.}~\bibnamefont{Fang}},
  \bibinfo{author}{\bibfnamefont{H.~C.} \bibnamefont{Po}},
  \bibinfo{author}{\bibfnamefont{A.}~\bibnamefont{Vishwanath}},
  \bibnamefont{and} \bibinfo{author}{\bibfnamefont{E.}~\bibnamefont{Kaxiras}},
  \bibinfo{journal}{arXiv:1907.06282}  (\bibinfo{year}{2019}{\natexlab{b}}).

\bibitem[{\citenamefont{Fang et~al.}(2019)\citenamefont{Fang, Carr, Zhu,
  Massatt, and Kaxiras}}]{Fang2019}
\bibinfo{author}{\bibfnamefont{S.}~\bibnamefont{Fang}},
  \bibinfo{author}{\bibfnamefont{S.}~\bibnamefont{Carr}},
  \bibinfo{author}{\bibfnamefont{Z.}~\bibnamefont{Zhu}},
  \bibinfo{author}{\bibfnamefont{D.}~\bibnamefont{Massatt}}, \bibnamefont{and}
  \bibinfo{author}{\bibfnamefont{E.}~\bibnamefont{Kaxiras}},
  \bibinfo{journal}{arXiv:1908.00058}  (\bibinfo{year}{2019}).

\bibitem[{\citenamefont{Yuan and Fu}(2018)}]{Yuan2018PRB}
\bibinfo{author}{\bibfnamefont{N.~F.~Q.} \bibnamefont{Yuan}} \bibnamefont{and}
  \bibinfo{author}{\bibfnamefont{L.}~\bibnamefont{Fu}},
  \bibinfo{journal}{Physical Review B} \textbf{\bibinfo{volume}{98}},
  \bibinfo{pages}{045103} (\bibinfo{year}{2018}).

\bibitem[{\citenamefont{Kang and Vafek}(2018)}]{Kang2018PRX}
\bibinfo{author}{\bibfnamefont{J.}~\bibnamefont{Kang}} \bibnamefont{and}
  \bibinfo{author}{\bibfnamefont{O.}~\bibnamefont{Vafek}},
  \bibinfo{journal}{Physical Review X} \textbf{\bibinfo{volume}{8}},
  \bibinfo{pages}{031088} (\bibinfo{year}{2018}).

\bibitem[{\citenamefont{Stewart}(2011)}]{Stewart2011}
\bibinfo{author}{\bibfnamefont{G.~R.} \bibnamefont{Stewart}},
  \bibinfo{journal}{Reviews of Modern Physics} \textbf{\bibinfo{volume}{83}},
  \bibinfo{pages}{1589} (\bibinfo{year}{2011}).

\bibitem[{\citenamefont{Hirschfeld et~al.}(2011)\citenamefont{Hirschfeld,
  Korshunov, and Mazin}}]{Hirschfeld2011}
\bibinfo{author}{\bibfnamefont{P.~J.} \bibnamefont{Hirschfeld}},
  \bibinfo{author}{\bibfnamefont{M.~M.} \bibnamefont{Korshunov}},
  \bibnamefont{and} \bibinfo{author}{\bibfnamefont{I.~I.} \bibnamefont{Mazin}},
  \bibinfo{journal}{Reports on Progress in Physics}
  \textbf{\bibinfo{volume}{74}}, \bibinfo{pages}{124508}
  (\bibinfo{year}{2011}), ISSN \bibinfo{issn}{0034-4885 1361-6633}.

\bibitem[{\citenamefont{Seo et~al.}(2008)\citenamefont{Seo, Bernevig, and
  Hu}}]{Seo2018}
\bibinfo{author}{\bibfnamefont{K.}~\bibnamefont{Seo}},
  \bibinfo{author}{\bibfnamefont{B.~A.} \bibnamefont{Bernevig}},
  \bibnamefont{and} \bibinfo{author}{\bibfnamefont{J.}~\bibnamefont{Hu}},
  \bibinfo{journal}{Physical Review Letters} \textbf{\bibinfo{volume}{101}},
  \bibinfo{pages}{206404} (\bibinfo{year}{2008}).

\bibitem[{\citenamefont{Platt et~al.}(2013)\citenamefont{Platt, Hanke, and
  Thomale}}]{Platt2013}
\bibinfo{author}{\bibfnamefont{C.}~\bibnamefont{Platt}},
  \bibinfo{author}{\bibfnamefont{W.}~\bibnamefont{Hanke}}, \bibnamefont{and}
  \bibinfo{author}{\bibfnamefont{R.}~\bibnamefont{Thomale}},
  \bibinfo{journal}{Advances in Physics} \textbf{\bibinfo{volume}{62}},
  \bibinfo{pages}{453} (\bibinfo{year}{2013}), ISSN \bibinfo{issn}{0001-8732}.

\bibitem[{\citenamefont{B?hm et~al.}(2018)\citenamefont{B?hm, Kretzschmar,
  Baum, Rehm, Jost, Hosseinian~Ahangharnejhad, Thomale, Platt, Maier, Hanke
  et~al.}}]{Boehm2018}
\bibinfo{author}{\bibfnamefont{T.}~\bibnamefont{B?hm}},
  \bibinfo{author}{\bibfnamefont{F.}~\bibnamefont{Kretzschmar}},
  \bibinfo{author}{\bibfnamefont{A.}~\bibnamefont{Baum}},
  \bibinfo{author}{\bibfnamefont{M.}~\bibnamefont{Rehm}},
  \bibinfo{author}{\bibfnamefont{D.}~\bibnamefont{Jost}},
  \bibinfo{author}{\bibfnamefont{R.}~\bibnamefont{Hosseinian~Ahangharnejhad}},
  \bibinfo{author}{\bibfnamefont{R.}~\bibnamefont{Thomale}},
  \bibinfo{author}{\bibfnamefont{C.}~\bibnamefont{Platt}},
  \bibinfo{author}{\bibfnamefont{T.~A.} \bibnamefont{Maier}},
  \bibinfo{author}{\bibfnamefont{W.}~\bibnamefont{Hanke}},
  \bibnamefont{et~al.}, \bibinfo{journal}{npj Quantum Materials}
  \textbf{\bibinfo{volume}{3}}, \bibinfo{pages}{48} (\bibinfo{year}{2018}),
  ISSN \bibinfo{issn}{2397-4648}.

\bibitem[{\citenamefont{Mackenzie and Maeno}(2003)}]{Mackenzie2003}
\bibinfo{author}{\bibfnamefont{A.~P.} \bibnamefont{Mackenzie}}
  \bibnamefont{and} \bibinfo{author}{\bibfnamefont{Y.}~\bibnamefont{Maeno}},
  \bibinfo{journal}{Reviews of Modern Physics} \textbf{\bibinfo{volume}{75}},
  \bibinfo{pages}{657} (\bibinfo{year}{2003}).

\bibitem[{\citenamefont{Wang et~al.}(2013)\citenamefont{Wang, Platt, Yang,
  Honerkamp, Zhang, Hanke, Rice, and Thomale}}]{WangQH2013}
\bibinfo{author}{\bibfnamefont{Q.~H.} \bibnamefont{Wang}},
  \bibinfo{author}{\bibfnamefont{C.}~\bibnamefont{Platt}},
  \bibinfo{author}{\bibfnamefont{Y.}~\bibnamefont{Yang}},
  \bibinfo{author}{\bibfnamefont{C.}~\bibnamefont{Honerkamp}},
  \bibinfo{author}{\bibfnamefont{F.~C.} \bibnamefont{Zhang}},
  \bibinfo{author}{\bibfnamefont{W.}~\bibnamefont{Hanke}},
  \bibinfo{author}{\bibfnamefont{T.~M.} \bibnamefont{Rice}}, \bibnamefont{and}
  \bibinfo{author}{\bibfnamefont{R.}~\bibnamefont{Thomale}},
  \bibinfo{journal}{EPL (Europhysics Letters)} \textbf{\bibinfo{volume}{104}},
  \bibinfo{pages}{17013} (\bibinfo{year}{2013}), ISSN \bibinfo{issn}{0295-5075
  1286-4854}.

\bibitem[{\citenamefont{Scaffidi and Simon}(2015)}]{Scaffidi2015}
\bibinfo{author}{\bibfnamefont{T.}~\bibnamefont{Scaffidi}} \bibnamefont{and}
  \bibinfo{author}{\bibfnamefont{S.~H.} \bibnamefont{Simon}},
  \bibinfo{journal}{Physical Review Letters} \textbf{\bibinfo{volume}{115}},
  \bibinfo{pages}{087003} (\bibinfo{year}{2015}).

\bibitem[{\citenamefont{Choi et~al.}(2019)\citenamefont{Choi, Kemmer, Peng,
  Thomson, Arora, Polski, Zhang, Ren, Alicea, Refael et~al.}}]{Choi2018}
\bibinfo{author}{\bibfnamefont{Y.}~\bibnamefont{Choi}},
  \bibinfo{author}{\bibfnamefont{J.}~\bibnamefont{Kemmer}},
  \bibinfo{author}{\bibfnamefont{Y.}~\bibnamefont{Peng}},
  \bibinfo{author}{\bibfnamefont{A.}~\bibnamefont{Thomson}},
  \bibinfo{author}{\bibfnamefont{H.}~\bibnamefont{Arora}},
  \bibinfo{author}{\bibfnamefont{R.}~\bibnamefont{Polski}},
  \bibinfo{author}{\bibfnamefont{Y.}~\bibnamefont{Zhang}},
  \bibinfo{author}{\bibfnamefont{H.}~\bibnamefont{Ren}},
  \bibinfo{author}{\bibfnamefont{J.}~\bibnamefont{Alicea}},
  \bibinfo{author}{\bibfnamefont{G.}~\bibnamefont{Refael}},
  \bibnamefont{et~al.}, \bibinfo{journal}{arXiv e-prints:1901.02997}
  (\bibinfo{year}{2019}).

\bibitem[{\citenamefont{Moon and Koshino}(2012)}]{Moon2012}
\bibinfo{author}{\bibfnamefont{P.}~\bibnamefont{Moon}} \bibnamefont{and}
  \bibinfo{author}{\bibfnamefont{M.}~\bibnamefont{Koshino}},
  \bibinfo{journal}{Physical Review B} \textbf{\bibinfo{volume}{85}},
  \bibinfo{pages}{195458} (\bibinfo{year}{2012}).

\bibitem[{\citenamefont{Lopes?dos?Santos
  et~al.}(2007)\citenamefont{Lopes?dos?Santos, Peres, and
  Castro?Neto}}]{LopesdosSantos2007}
\bibinfo{author}{\bibfnamefont{J.~M.~B.} \bibnamefont{Lopes?dos?Santos}},
  \bibinfo{author}{\bibfnamefont{N.~M.~R.} \bibnamefont{Peres}},
  \bibnamefont{and} \bibinfo{author}{\bibfnamefont{A.~H.}
  \bibnamefont{Castro?Neto}}, \bibinfo{journal}{Physical Review Letters}
  \textbf{\bibinfo{volume}{99}}, \bibinfo{pages}{256802}
  (\bibinfo{year}{2007}).

\bibitem[{\citenamefont{Bistritzer and MacDonald}(2011)}]{Bistritzer2011}
\bibinfo{author}{\bibfnamefont{R.}~\bibnamefont{Bistritzer}} \bibnamefont{and}
  \bibinfo{author}{\bibfnamefont{A.~H.} \bibnamefont{MacDonald}},
  \bibinfo{journal}{Proceedings of the National Academy of Sciences}
  \textbf{\bibinfo{volume}{108}}, \bibinfo{pages}{12233}
  (\bibinfo{year}{2011}).

\bibitem[{\citenamefont{Tomarken et~al.}(2019)\citenamefont{Tomarken, Cao,
  Demir, Watanabe, Taniguchi, Jarillo-Herrero, and Ashoori}}]{Tomarken2019}
\bibinfo{author}{\bibfnamefont{S.~L.} \bibnamefont{Tomarken}},
  \bibinfo{author}{\bibfnamefont{Y.}~\bibnamefont{Cao}},
  \bibinfo{author}{\bibfnamefont{A.}~\bibnamefont{Demir}},
  \bibinfo{author}{\bibfnamefont{K.}~\bibnamefont{Watanabe}},
  \bibinfo{author}{\bibfnamefont{T.}~\bibnamefont{Taniguchi}},
  \bibinfo{author}{\bibfnamefont{P.}~\bibnamefont{Jarillo-Herrero}},
  \bibnamefont{and} \bibinfo{author}{\bibfnamefont{R.~C.}
  \bibnamefont{Ashoori}}, \bibinfo{journal}{Physical Review Letters}
  \textbf{\bibinfo{volume}{123}}, \bibinfo{pages}{046601}
  (\bibinfo{year}{2019}).

\bibitem[{\citenamefont{Hejazi et~al.}(2019)\citenamefont{Hejazi, Liu,
  Shapourian, Chen, and Balents}}]{Hejazi2019}
\bibinfo{author}{\bibfnamefont{K.}~\bibnamefont{Hejazi}},
  \bibinfo{author}{\bibfnamefont{C.}~\bibnamefont{Liu}},
  \bibinfo{author}{\bibfnamefont{H.}~\bibnamefont{Shapourian}},
  \bibinfo{author}{\bibfnamefont{X.}~\bibnamefont{Chen}}, \bibnamefont{and}
  \bibinfo{author}{\bibfnamefont{L.}~\bibnamefont{Balents}},
  \bibinfo{journal}{Physical Review B} \textbf{\bibinfo{volume}{99}},
  \bibinfo{pages}{035111} (\bibinfo{year}{2019}).

\bibitem[{\citenamefont{{Pizarro} et~al.}(2019)\citenamefont{{Pizarro},
  {R{\"o}sner}, {Thomale}, {Valent{\'\i}}, and {Wehling}}}]{Pizarro2019}
\bibinfo{author}{\bibfnamefont{J.~M.} \bibnamefont{{Pizarro}}},
  \bibinfo{author}{\bibfnamefont{M.}~\bibnamefont{{R{\"o}sner}}},
  \bibinfo{author}{\bibfnamefont{R.}~\bibnamefont{{Thomale}}},
  \bibinfo{author}{\bibfnamefont{R.}~\bibnamefont{{Valent{\'\i}}}},
  \bibnamefont{and} \bibinfo{author}{\bibfnamefont{T.~O.}
  \bibnamefont{{Wehling}}}, \bibinfo{journal}{arXiv:1904.11765}
  (\bibinfo{year}{2019}).

\bibitem[{\citenamefont{Pruschke and Bulla}(2005)}]{Pruschke2005}
\bibinfo{author}{\bibfnamefont{T.}~\bibnamefont{Pruschke}} \bibnamefont{and}
  \bibinfo{author}{\bibfnamefont{R.}~\bibnamefont{Bulla}},
  \bibinfo{journal}{Eur. Phys. J. B} \textbf{\bibinfo{volume}{44}},
  \bibinfo{pages}{217} (\bibinfo{year}{2005}).

\bibitem[{\citenamefont{Haule and Kotliar}(2009)}]{Haule2009}
\bibinfo{author}{\bibfnamefont{K.}~\bibnamefont{Haule}} \bibnamefont{and}
  \bibinfo{author}{\bibfnamefont{G.}~\bibnamefont{Kotliar}},
  \bibinfo{journal}{New Journal of Physics} \textbf{\bibinfo{volume}{11}},
  \bibinfo{pages}{025021} (\bibinfo{year}{2009}), ISSN
  \bibinfo{issn}{1367-2630}.

\bibitem[{\citenamefont{Kohn and Luttinger}(1965)}]{Kohn1965}
\bibinfo{author}{\bibfnamefont{W.}~\bibnamefont{Kohn}} \bibnamefont{and}
  \bibinfo{author}{\bibfnamefont{J.~M.} \bibnamefont{Luttinger}},
  \bibinfo{journal}{Physical Review Letters} \textbf{\bibinfo{volume}{15}},
  \bibinfo{pages}{524} (\bibinfo{year}{1965}).

\bibitem[{\citenamefont{Metzner et~al.}(2012)\citenamefont{Metzner, Salmhofer,
  Honerkamp, Meden, and Sch?nhammer}}]{Metzner2012}
\bibinfo{author}{\bibfnamefont{W.}~\bibnamefont{Metzner}},
  \bibinfo{author}{\bibfnamefont{M.}~\bibnamefont{Salmhofer}},
  \bibinfo{author}{\bibfnamefont{C.}~\bibnamefont{Honerkamp}},
  \bibinfo{author}{\bibfnamefont{V.}~\bibnamefont{Meden}}, \bibnamefont{and}
  \bibinfo{author}{\bibfnamefont{K.}~\bibnamefont{Sch?nhammer}},
  \bibinfo{journal}{Reviews of Modern Physics} \textbf{\bibinfo{volume}{84}},
  \bibinfo{pages}{299} (\bibinfo{year}{2012}).

\bibitem[{\citenamefont{Kohn}(1973)}]{Kohn1973}
\bibinfo{author}{\bibfnamefont{W.}~\bibnamefont{Kohn}},
  \bibinfo{journal}{Physical Review B} \textbf{\bibinfo{volume}{7}},
  \bibinfo{pages}{4388} (\bibinfo{year}{1973}).

\bibitem[{\citenamefont{Cloizeaux}(1964)}]{Cloizeaux1964}
\bibinfo{author}{\bibfnamefont{J.~D.} \bibnamefont{Cloizeaux}},
  \bibinfo{journal}{Physical Review} \textbf{\bibinfo{volume}{135}},
  \bibinfo{pages}{A698} (\bibinfo{year}{1964}).

\bibitem[{\citenamefont{Elster et~al.}(2015)\citenamefont{Elster, Platt,
  Thomale, Hanke, and Hankiewicz}}]{Elster2015}
\bibinfo{author}{\bibfnamefont{L.}~\bibnamefont{Elster}},
  \bibinfo{author}{\bibfnamefont{C.}~\bibnamefont{Platt}},
  \bibinfo{author}{\bibfnamefont{R.}~\bibnamefont{Thomale}},
  \bibinfo{author}{\bibfnamefont{W.}~\bibnamefont{Hanke}}, \bibnamefont{and}
  \bibinfo{author}{\bibfnamefont{E.~M.} \bibnamefont{Hankiewicz}},
  \bibinfo{journal}{Nature Communications} \textbf{\bibinfo{volume}{6}},
  \bibinfo{pages}{8232} (\bibinfo{year}{2015}).

\bibitem[{\citenamefont{Yang et~al.}(2019)\citenamefont{Yang, Gao, and
  Fa}}]{YangH2019}
\bibinfo{author}{\bibfnamefont{H.}~\bibnamefont{Yang}},
  \bibinfo{author}{\bibfnamefont{Z.}~\bibnamefont{Gao}}, \bibnamefont{and}
  \bibinfo{author}{\bibfnamefont{W.}~\bibnamefont{Fa}},
  \bibinfo{journal}{arXiv:1908.09555}  (\bibinfo{year}{2019}).

\bibitem[{\citenamefont{Kemper et~al.}(2010)\citenamefont{Kemper, Maier,
  Graser, Cheng, Hirschfeld, and Scalapino}}]{Kemper2010}
\bibinfo{author}{\bibfnamefont{A.~F.} \bibnamefont{Kemper}},
  \bibinfo{author}{\bibfnamefont{T.~A.} \bibnamefont{Maier}},
  \bibinfo{author}{\bibfnamefont{S.}~\bibnamefont{Graser}},
  \bibinfo{author}{\bibfnamefont{H.~P.} \bibnamefont{Cheng}},
  \bibinfo{author}{\bibfnamefont{P.~J.} \bibnamefont{Hirschfeld}},
  \bibnamefont{and} \bibinfo{author}{\bibfnamefont{D.~J.}
  \bibnamefont{Scalapino}}, \bibinfo{journal}{New Journal of Physics}
  \textbf{\bibinfo{volume}{12}}, \bibinfo{pages}{073030}
  (\bibinfo{year}{2010}), ISSN \bibinfo{issn}{1367-2630}.

\bibitem[{\citenamefont{Bickers et~al.}(1989)\citenamefont{Bickers, Scalapino,
  and White}}]{Bickers1989}
\bibinfo{author}{\bibfnamefont{N.~E.} \bibnamefont{Bickers}},
  \bibinfo{author}{\bibfnamefont{D.~J.} \bibnamefont{Scalapino}},
  \bibnamefont{and} \bibinfo{author}{\bibfnamefont{S.~R.} \bibnamefont{White}},
  \bibinfo{journal}{Physical Review Letters} \textbf{\bibinfo{volume}{62}},
  \bibinfo{pages}{961} (\bibinfo{year}{1989}).

\bibitem[{\citenamefont{Kubo}(2007)}]{Kubo2007}
\bibinfo{author}{\bibfnamefont{K.}~\bibnamefont{Kubo}},
  \bibinfo{journal}{Physical Review B} \textbf{\bibinfo{volume}{75}},
  \bibinfo{pages}{224509} (\bibinfo{year}{2007}).

\bibitem[{\citenamefont{Wu et~al.}(2014)\citenamefont{Wu, Yuan, Liang, Fan, and
  Hu}}]{Wu2014}
\bibinfo{author}{\bibfnamefont{X.}~\bibnamefont{Wu}},
  \bibinfo{author}{\bibfnamefont{J.}~\bibnamefont{Yuan}},
  \bibinfo{author}{\bibfnamefont{Y.}~\bibnamefont{Liang}},
  \bibinfo{author}{\bibfnamefont{H.}~\bibnamefont{Fan}}, \bibnamefont{and}
  \bibinfo{author}{\bibfnamefont{J.}~\bibnamefont{Hu}}, \bibinfo{journal}{EPL
  (Europhysics Letters)} \textbf{\bibinfo{volume}{108}}, \bibinfo{pages}{27006}
  (\bibinfo{year}{2014}).

\bibitem[{\citenamefont{Wu et~al.}(2015)\citenamefont{Wu, Yang, Le, Fan, and
  Hu}}]{Wu2015}
\bibinfo{author}{\bibfnamefont{X.}~\bibnamefont{Wu}},
  \bibinfo{author}{\bibfnamefont{F.}~\bibnamefont{Yang}},
  \bibinfo{author}{\bibfnamefont{C.}~\bibnamefont{Le}},
  \bibinfo{author}{\bibfnamefont{H.}~\bibnamefont{Fan}}, \bibnamefont{and}
  \bibinfo{author}{\bibfnamefont{J.}~\bibnamefont{Hu}},
  \bibinfo{journal}{Physical Review B} \textbf{\bibinfo{volume}{92}},
  \bibinfo{pages}{104511} (\bibinfo{year}{2015}).

\bibitem[{\citenamefont{Honerkamp}(2008)}]{Honerkamp2008}
\bibinfo{author}{\bibfnamefont{C.}~\bibnamefont{Honerkamp}},
  \bibinfo{journal}{Physical Review Letters} \textbf{\bibinfo{volume}{100}},
  \bibinfo{pages}{146404} (\bibinfo{year}{2008}).


  \bibitem[{\citenamefont{Scalapino}(1995)}]{Scalapino1995}
\bibinfo{author}{\bibfnamefont{D.}~\bibnamefont{Scalapino}},
  \bibinfo{journal}{Physics Reports} \textbf{\bibinfo{volume}{250}},
  \bibinfo{pages}{329} (\bibinfo{year}{1995}).

\end{thebibliography}

\end{document}